\documentclass[english,ruled,vlined]{IEEEtran}
\usepackage[T1]{fontenc}
\usepackage[latin9]{inputenc}
\usepackage{color}
\usepackage{array}
\usepackage{units}
\usepackage{multirow}
\usepackage{algorithm2e}
\usepackage{amsmath}
\usepackage{amssymb}
\usepackage{stackrel}
\usepackage{graphicx}
\usepackage{wasysym}
\usepackage{esint}

\makeatletter

\providecommand{\tabularnewline}{\\}

\usepackage{subfigure}
\usepackage{cite}
\usepackage{blindtext}

\usepackage{algpseudocode}
\usepackage[justification=centering]{caption}

\makeatother

\usepackage{babel}
\begin{document}

\title{Image Moment Models for Extended Object Tracking}

\author{\author{Gang~Yao,~Ashwin~Dani \thanks{Gang~Yao~and~Ashwin~Dani are with the Department of Electrical and Computer Engineering, 
University of Connecticut Storrs, Connecticut, USA. e-mail: gang.yao@uconn.edu, ashwin.dani@uconn.edu  }}}
\maketitle
\begin{abstract}
In this paper, a novel image moments based model for shape estimation
and tracking of an object moving with a complex trajectory is presented.
The camera is assumed to be stationary looking at a moving object.
Point features inside the object are sampled as measurements. An ellipsoidal
approximation of the shape is assumed as a primitive shape. The shape
of an ellipse is estimated using a combination of image moments. Dynamic
model of image moments when the object moves under the constant velocity
or coordinated turn motion model is derived as a function for the
shape estimation of the object. An Unscented Kalman Filter-Interacting
Multiple Model (UKF-IMM) filter algorithm is applied to estimate the
shape of the object (approximated as an ellipse) and track its position
and velocity. A likelihood function based on average log-likelihood
is derived for the IMM filter. Simulation results of the proposed
UKF-IMM algorithm with the image moments based models are presented
that show the estimations of the shape of the object moving in complex
trajectories. Comparison results, using intersection over union (IOU),
and position and velocity root mean square errors (RMSE) as metrics,
with a benchmark algorithm from literature are presented. Results
on real image data captured from the quadcopter are also presented.
\end{abstract}

\begin{IEEEkeywords}
Extended Object Tracking, Shape Estimation, Image Moments Dynamic
Model, Log-Likelihood for filtering
\end{IEEEkeywords}

\section{\textcolor{black}{Introduction}}

Traditional target tracking literature\cite{BarShalom2011}, such
as simultaneous localization and mapping (SLAM) \cite{dani2013image,yang2017vision},
structure from motion (SfM) \cite{dani2012single,chwa2016range} and
target tracking \cite{yao2016gyro}, models the targets as point targets.
Once the estimation is performed, another layer of optimization is
used to estimate the shape of the target. With the increased resolution
of the modern sensors, such as phased array radar, laser range finder,
2D/3D cameras, the sensors are capable of giving more than one point
measurement from an observed target at a single time instance. For
instance, in a camera image, multiple SIFT/SURF points can be obtained
inside a chosen region of interest (ROI) or 3D cameras, such as Kinect
camera gives a collection of points in a given ROI. The multiple measurements
from a target can be used to estimate and track not only the position
and velocity of the centroid but also its spatial extent. The combined
target tracking and shape estimation is commonly referred to as an
extended object tracking (EOT) problem \textcolor{black}{\cite{granstrom2016extended,mihaylova2014overview}}. 

Multiple feature points such as SIFT and SURF points can not be identified
consistently and tracked individually over long period of time inside
an object or multiple objects. With multiple noisy measure points
generated from the target at each time step without association, the
target can be roughly estimated as an ellipse shape, which will provide
the kinematic (position and velocity of the centroid) and spatial
extent information (orientation and size) useful for real world applications.
The extended object is modeled by stick \cite{vermaak2005sequential},
Gaussian mixture model \cite{granstrom2010gaussian}, rectangle \cite{granstrom2014multiple},
Gaussian process model \cite{wahlstrom2015extended}, and splines
\cite{zea2016tracking}. Two widely used models to represent targets
with spatial extent are random matrix model (RMM) \cite{koch2008bayesian}
and elliptic random hyper-surface model (RHM) \cite{baum2010extended},
where the true shape of the object is approximated by an ellipse.
In RMM, the shape of the target object is represented by using a symmetric
positive definite (SPD) matrix. The elements of the matrix along with
the centroid of the object are used as a state vector, which is estimated
by using a filter. Multiple improvements to the RMM model are presented
in literature \cite{feldmann2009advances,lan2014tracking,Granstroem2014,lan2016tracking}.
The situation when the measurement noise is comparable to the extent
of the target and can not be neglected is considered in \cite{feldmann2009advances,feldmann2011tracking}.
Considering the target will change the size and shape abruptly especially
during the maneuvering movement, the rotation matrix or scaling matrix
is multiplied on both sides of the positive symmetric matrix and the
corresponding filters are derived in \cite{lan2014tracking,Granstroem2014,lan2016tracking}.
The RHM model assumes each measurement source lies on a scaled version
of the true ellipse describing the object, and the extent of the object
is represented by the entries from the Cholesky decomposition of the
SPD matrix \cite{baum2011shape,baum2012tracking,baum2014extended}.
In \cite{yang2016second}, a multiplicative noise term in the measurement
equation is used to model the spatial distribution of the measurements
and a second order extended Kalman filter is derived for a closed
form recursive measurement update. In \cite{baum2010extendedcompare},
comparisons between the RHM with RMM are illustrated. RHM with Fourier
series expansion and level-set are applied for modeling star-convex
and non-convex shapes, respectively \cite{baum2014extended,zea2016level}.
By approximating the complex shapes as the combination of multiple
ellipse sub-objects, the elliptic RMMs are investigated to model irregular
shapes \cite{lan2014tracking,zong2016improved}. A comprehensive overview
of the extended object tracking can be found in \cite{mihaylova2014overview,granstrom2016extended}.

The dynamic model for a moving extended object describes how the target's
kinematic parameters and extent evolve over time. For tracking a point
object, the kinematic parameters such as position, velocity or acceleration
can fully describe the state of the object. However, for an extended
object, the object shape estimation is also important, especially
when the target conducts maneuvering motion or the shape of the extended
target changes abruptly. For tracking extended object using RMM, there
is no explicit dynamic model and the update for the extent is based
on simple heuristics which increase the extent's covariance, while
keeping the expected value constant \cite{koch2008bayesian}. An alternative
to the heuristic update is to use Wishart distribution to approximate
the transition density of the spatial extent \cite{koch2008bayesian,lan2012trackingA,Granstroem2014}.
The prediction update of extended targets within the RMM framework
is explored by multiplying the rotation matrix or scaling matrix on
the both sides of the positive symmetric matrix in \cite{lan2016tracking,Granstroem2014}.
In \cite{Granstroem2014}, a comprehensive comparison results between
four process models are presented. For tracking elliptic extended
object using RHM, the covariance matrix of the uncertainty of the
object's shape parameters is increased at each time step to capture
the variations in the shape \cite{baum2010extended}.

Image moments have found a wide use in tracking, visual servoing and
pattern recognition \cite{chaumette2004image,tahri2005point,dani2013image,spica2015plane}.
Hu's moments \cite{hu1962visual} invariant under translation, rotation
and scaling of the object, are widely investigated in pattern recognition.
In this paper, \textcolor{black}{an alternative representation, using
image moments, to describe an ellipse shape that can be used to approximate
an extended object is presented. }Dynamic models of image moments
that are used to represent an extended object for the target moving
in an uniform motion and a coordinated turn motion are presented.
The image moments based RHM is used with the interacting multiple
model (IMM) approach \cite{bar2004estimation,kirubarajan2003kalman,granstrom2015systematic}
for tracking extended target undergoing complex trajectories. A novel
likelihood function based on average log-likelihood is derived for
the IMM. An unscented Kalman filter (UFK) is used to estimate the
states of each individual model of the UKF-IMM filter. The UKF-IMM
approach assumes the target obeys one of the finite number of motion
models and identifies the beginning and the end of the motion models
by updating the mode probabilities. The adaptation via model probability
update of the UKF-IMM approach keeps the estimation errors low, both
during maneuvers as well as non-maneuver intervals. The contributions
of the paper are briefly summarized as follows:
\begin{itemize}
\item \textcolor{black}{The} minimal, complete, and non-ambiguous representation
\textcolor{black}{of an elliptic object based on image moments is
presented for extended object tracking. }UKF-IMM filter is adopted
based on the multiple dynamic models and corresponding image moments
based RHM.
\item A novel method of calculating the likelihood function, based on average
log-likelihood of the image moments based RHM, is proposed for the
UKF-IMM filter. In order to estimate the model probability consistently,
the calculation of the average log-likelihood function by unscented
transformation is proposed.
\item Results of the UKF-IMM filter with the image moments based model are
presented and compared with a benchmark algorithm to validate the
performance of the proposed approach.
\end{itemize}
Rest of the paper is organized as follows. In Section \ref{sec:Image-Moment-based-RHM},
the image moments based random hypersurface model is proposed to approximate
an elliptic object and its dynamic models are analytically derived.
Following the framework of the random hypersurface model, the measurement
model is also provided. in Section \ref{sec:Unscented-Kalman-Filter},
the Bayesian inference of the position, velocity and extent of the
object from the noisy measurement points uniformly generated from
the object is illustrated. Since the dynamic or measurement model
is nonlinear, UKF is applied to estimate the extended object. For
tracking the moving target switching between maneuvering and non-maneuvering
motions, the proposed image moments based RHM is embedded within the
framework of the interacting multiple model (IMM) in section \ref{sec:Tracking-extended-target with IMM}.
The UKF-IMM algorithm is illustrated with the proposed image moments
based RHM, and the algorithm for the calculation of the likelihood
function by using the average log-likelihood function and unscented
transformation is also proposed. In section \ref{sec:Evaluation},
the proposed image moments based RHM with its dynamic models is evaluated
in three tests: (1) static scenario for validating the measurement
model; (2) constant velocity and coordinated turn motion to validate
the dynamic models; (3) two complex trajectories are used to validate
the UKF-IMM algorithm with the proposed image moments based RHM, and
its performance is compared with the RMM models in \cite{feldmann2011tracking}
as the benchmark. The estimation results show that the proposed model
provides comparable and accurate results. In Section \ref{sec:Experiment},
the proposed algorithm is applied for tracking a moving car with the
real trajectory. Conclusion and future work are given in Section \ref{sec:Conclusion-and-Future}.
To improve legibility, the subindices, such as the time step $k$
and the measurement number $l$ will be dropped unless needed in the
following.

\section{Image Moments based Random Hypersurface Model\label{sec:Image-Moment-based-RHM}}

\subsection{Representation of the Ellipse using Image Moments}

In this section, a generalized representation of the ellipse using
image moments is presented. The $(i+j\mathrm{)th}$ moment of an object
$m_{ij}$ in a 2D plane is defined by \cite{chaumette2004image}

\begin{equation}
m_{ij}=\iintop_{R(t)}x^{i}y^{j}dxdy,\;\:\forall\,i,j\in\mathbb{N}
\end{equation}
where $R(t)$ is the surface of the object and $\mathbb{N}$ is a
set of natural numbers. The centered moment is defined as \cite{chaumette2004image}

\begin{equation}
\eta_{ij}=\iintop_{R(t)}h(\bar{x},\bar{y})dxdy
\end{equation}
where $h(\bar{x},\bar{y})=(\bar{x})^{i}(\bar{y})^{j}$, $\bar{x}=x-x_{c}$,
$\bar{y}=y-y_{c}$ and $(x_{c,}y_{c})$ is the centroid of the object.

Any point on the surface of the object can be represented as a point
located on the boundary of the scaled ellipse. The general equation
of a family of ellipses in terms of semi-major, and semi-minor axes,
centroid, and orientation is given by

\begin{equation}
\frac{(x-x_{c}+t(y-y_{c}))^{2}}{a_{1}^{2}(1+t^{2})}+\frac{(y-y_{c}-t(x-x_{c}))^{2}}{a_{2}^{2}(1+t^{2})}-s^{2}=0\label{eq:ellipse}
\end{equation}
where $a_{1}$ and $a_{2}$ are its semi-major and semi-minor axes,
respectively, $t$ is related to the orientation of ellipse $\alpha$,
as $t=\tan\alpha$, and $s$ is a scale factor. The points $(x,y)$
inside the ellipse can be represented by varying $s$ from $0$ to
$1$ in (\ref{eq:ellipse}). Rewriting (\ref{eq:ellipse}) as follows

\begin{equation}
\frac{a_{1}^{2}t^{2}+a_{2}^{2}}{a_{1}^{2}a_{2}^{2}(1+t^{2})}\bar{x}^{2}+\frac{t^{2}a_{2}^{2}+a_{1}^{2}}{a_{1}^{2}a_{2}^{2}(1+t^{2})}\bar{y}^{2}+\frac{a_{2}^{2}-a_{1}^{2}}{a_{1}^{2}a_{2}^{2}}\frac{2t}{1+t^{2}}\bar{x}\bar{y}=s^{2}\label{eq:ellipse1}
\end{equation}
Consider normalized centered moments $n_{11}=\frac{\eta_{11}}{a}$,
$n_{02}=\frac{\eta_{02}}{a}$, $n_{20}=\frac{\eta_{20}}{a}$, where
$a$ is the area of the ellipse, $\eta_{11}$, $\eta_{02}$, and $\eta_{20}$
are centered moments. The following relationships between parameters
of ellipse $a_{1},$ $a_{2}$, $t$, and the normalized centered image
moments ($n_{20}$, $n_{02}$, $n_{11}$) can be derived \cite{chaumette2004image} 

\begin{align}
a_{1}^{2} & =2\left(n_{02}+n_{20}+\sqrt{\left(n_{20}-n_{02}\right)^{2}+4n_{11}^{2}}\right)\nonumber \\
a_{2}^{2} & =2\left(n_{02}+n_{20}-\sqrt{\left(n_{20}-n_{02}\right)^{2}+4n_{11}^{2}}\right)\label{eq:InverseParameters}\\
t & =\frac{1}{2n_{11}}\left(n_{02}-n_{20}+\sqrt{\left(n_{20}-n_{02}\right)^{2}+4n_{11}^{2}}\right)\nonumber 
\end{align}
Substituting (\ref{eq:InverseParameters}) into (\ref{eq:ellipse1}),
the following expression is obtained

\begin{equation}
\frac{4n_{02}}{a_{1}^{2}a_{2}^{2}}\bar{x}^{2}+\frac{4n_{20}}{a_{1}^{2}a_{2}^{2}}\bar{y}^{2}-\frac{8n_{11}}{a_{1}^{2}a_{2}^{2}}\bar{x}\bar{y}=s^{2}\label{eq:7-1}
\end{equation}
The area of ellipse, $a$, can be written in normalized centered moments
$n_{ij}$ and parameters $a_{1}$, and $a_{2}$ as follows \cite{chaumette2004image}

\begin{equation}
a=\pi a_{1}a_{2}=4\pi\sqrt{n_{20}n_{02}-n_{11}^{2}}\label{eq:8-1}
\end{equation}
Using (\ref{eq:8-1}), (\ref{eq:7-1}) can be represented as follows

\begin{equation}
\begin{split}\frac{n_{02}}{4\left(n_{20}n_{02}-n_{11}^{2}\right)}\bar{x}^{2}+\frac{n_{20}}{4\left(n_{20}n_{02}-n_{11}^{2}\right)}\bar{y}^{2}\\
-\frac{2n_{11}}{4\left(n_{20}n_{02}-n_{11}^{2}\right)}\bar{x}\bar{y}=s^{2}
\end{split}
\label{eq:9-1}
\end{equation}
Let $\mathbf{p}=[\mathbf{p}_{\mathrm{IM}}^{T},\mathbf{p}_{\mathrm{pos}}^{T}]^{T}$,
where $\mathbf{p}_{\mathrm{IM}}=[n_{11},n_{20},n_{02}]^{T}$ can be
used to estimate the shape of the ellipse and $\mathbf{p}_{\mathrm{pos}}=[x_{c},y_{c}]^{T}$
represents the location of the centroid of the ellipse. An ellipse
can be expressed using minimal, complete, and non-ambiguous representation
of parameters $\mathbf{p}$, in the following form
\begin{align}
g(x,y,\mathbf{p})\!\!= & \frac{n_{02}}{4\left(n_{20}n_{02}-n_{11}^{2}\right)}\bar{x}^{2}+\frac{n_{20}}{4\left(n_{20}n_{02}-n_{11}^{2}\right)}\bar{y}^{2}\label{eq:moment_ellipse_cartesian}\\
 & -\frac{2n_{11}}{4\left(n_{20}n_{02}-n_{11}^{2}\right)}\bar{x}\bar{y}-s^{2}=0\nonumber 
\end{align}

\subsection{Dynamic Motion Models\label{subsec:Dynamic-model}}

In order to derive the differential equation for $n_{ij}$, the time
derivative of the centered moment, $\eta_{ij}$ is derived first.
The time derivative of centered moment $\eta_{ij}$ can be obtained
from the time derivative of the contour of the ellipse as \cite{chaumette2004image}

\begin{equation}
\dot{\eta}_{ij}=\ointop_{C(t)}h(x,y)\mathbf{v}^{T}\bar{\mathbf{n}}dl\label{eq:dynamic of momnets}
\end{equation}
where $C(t)$ is the contour of the ellipse, $\mathbf{v}=\left[\dot{x},\dot{y}\right]{}^{T}$
is the velocity of the contour point $\mathbf{x}=\left[x,y\right]^{T}$,
$\mathbf{\bar{n}}$ is the unitary vector normal to $C(t)$ at point
$\mathbf{x}$, and $dl$ is an infinitesimal element of $C(t)$. If
$C(t)$ is piece-wise continuous, and vector $h(x,y)\mathbf{\dot{x}}$
is tangent to $R(t)$ and continuously differentiable, $\forall\mathbf{x}\in R(t)$,
the Green's theorem can be used to represent (\ref{eq:dynamic of momnets})
as \cite{chaumette2004image}

\begin{equation}
\dot{\eta}_{ij}=\iint_{R(t)}\mathrm{div}[h(x,y)\mathbf{v}]dxdy\label{eq:dynamic of div}
\end{equation}
Using the constant velocity and coordinated turn models, specific
differential equation of $\eta_{ij}$ is derived for each case.

\subsubsection{Linear Motion Model}

When an elliptical object is moving with a linear motion, each point
inside the ellipse at time $t$ obeys $\mathbf{v}=\mathbf{v}_{0}+\mathbf{a}t$,
where $\mathbf{v_{0}}\in\mathbb{R}^{2}$ is the initial velocity and
$\mathbf{a}\in\mathbb{R}^{2}$ is the acceleration. The centered moments
of the ellipse $\mu_{ij}$ can be calculated by putting $h(x,y)=(x-x_{c})^{i}(y-y_{c})^{j}$
in (\ref{eq:dynamic of div}) as
\begin{equation}
\begin{aligned}\dot{\eta}_{ij}= & \iint_{R(t)}[\frac{\partial h}{\partial x}\dot{x}+\frac{\partial h}{\partial y}\dot{y}+h(x,y)(\frac{\partial\dot{x}}{\partial x}+\frac{\partial\dot{y}}{\partial y})]dxdy\end{aligned}
\label{eq:12}
\end{equation}
Since $\frac{\partial h}{\partial x}$ and $\frac{\partial h}{\partial y}$
are odd functions and $R$ is symmetric with respect to the centroid,
the state space representation of the normalized centered moments
of the ellipse $\mathbf{p}_{\mathrm{IM}}=\left[n_{11},\:n_{20},\:n_{02}\right]^{T}$
is
\begin{equation}
\mathbf{\dot{p}}_{\mathrm{IM}}=0\label{eq:linear motion}
\end{equation}

The state at discrete time $k$ is given by $\mathbf{p}_{k}=[\begin{array}{cc}
\mathbf{p}_{\mathit{\mathrm{IM}},k}^{T}, & \mathbf{p}_{\mathrm{CV},k}^{T}\end{array}]^{T}$, where $\mathbf{p}_{\mathit{\mathrm{IM}},k}$ is a component of the
state related to image moments, $\mathbf{p}_{\mathrm{CV},k}=[\begin{array}{cccc}
x_{c,k}, & \dot{x}_{c,k}, & y_{c,k}, & \dot{y}_{c,k}\end{array}]^{T}$ is the vector that includes the position and velocity of the centroid
of the extended object. The discretized state equation is given as
follows

\begin{equation}
\mathbf{p}_{k+1}=\mathbf{F}_{\mathrm{CV}}\mathbf{p}_{k}+\mathbf{w}_{k}\label{eq:cvm-1}
\end{equation}
where the state transition matrix $\mathbf{F}_{\mathrm{CV}}=\mathbf{diag}(\mathbf{I}_{3\times3},\mathbf{A}),$
with $\mathbf{A} = \left[\begin{smallmatrix} 1&T&0&0 \\ 0&1&0&0 \\ 0&0&1&T \\ 0&0&0&1 \end{smallmatrix} \right]$,
and $\mathbf{w}_{k}$ is the zero-mean Gaussian noise with covariance
matrix $\mathbf{C}_{\mathrm{CV},k}=\mathbf{diag}(\mathbf{C}_{\mathrm{IM},k},\mathbf{C}_{k}^{\mathbf{w}},\mathbf{C}_{k}^{\mathbf{w}})$,
where $\mathbf{C}_{\mathrm{IM},k}\in\mathbb{R}^{3\times3}$ is the
noise covariance for the image moments and $\mathbf{C}_k^\mathbf{w} = \left[\begin{smallmatrix} \frac{1}{3}T^3&\frac{1}{2}T^2 \\ \frac{1}{2}T^2&T  \end{smallmatrix} \right]q$
, where $q$ is the power spectral density. Notice that the discretized
white noise acceleration model is adopted for the state vector $\mathbf{p}_{\mathrm{CV},k}^{T}$,
which is the same as the dynamic model for point based tracking. Other
kinematic models for point based tracking also can be used for the
state vector $\mathbf{p}_{\mathrm{CV},k}^{T}$ and can be found in
\cite{bar2004estimation}.

\subsubsection{Coordinated Turn Motion Model}

\begin{figure}
\centering{}\includegraphics[width=0.95\columnwidth]{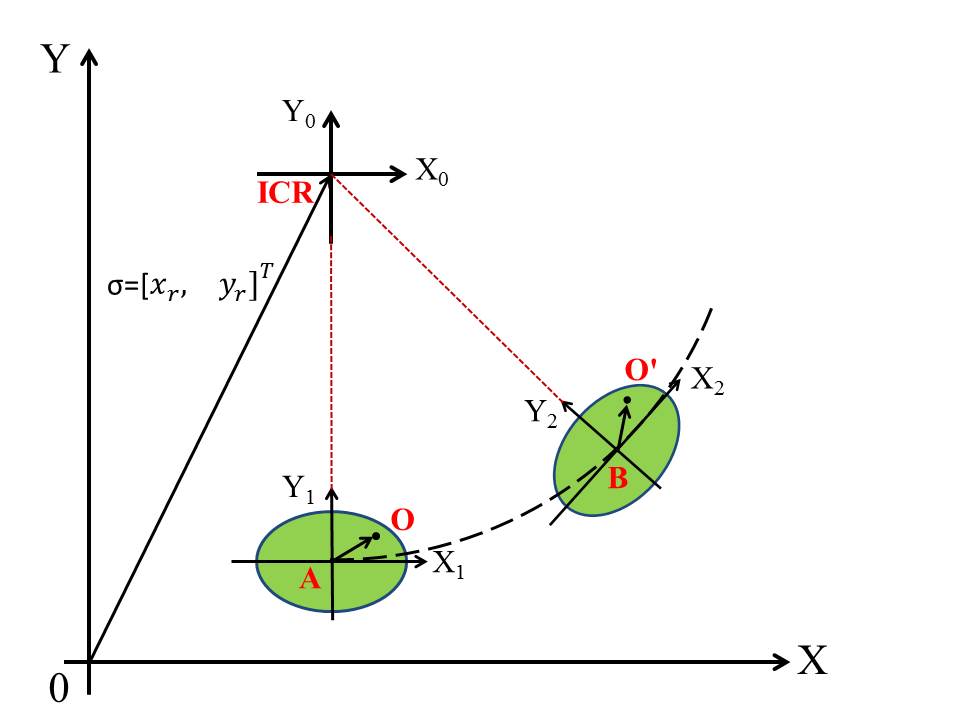}\caption{Coordinated turn model of the elliptic extended object.\label{fig:Coordinate-turning-ellipse}}
\end{figure}

Coordinated turn (CT) model, characterized by constant turning rate
and constant speed, are commonly used in tracking applications (cf.
\cite{bar2004estimation}). An elliptic extended object during the
coordinated turn is shown in Fig. \ref{fig:Coordinate-turning-ellipse}.
For any point $O(x,y)$ that belongs to the ellipse moving with a
CT motion, the motion model of the ellipse can be represented as follows 

\begin{equation}
\begin{aligned}\dot{x}= & -\omega(y-y_{r})\\
\dot{y}= & \omega(x-x_{r})
\end{aligned}
\label{eq:14}
\end{equation}
where $\omega$ is the turning rate and $\sigma=\left[x_{r},y_{r}\right]^{T}$
is the displacement between the origins of the reference frame $XY$
and reference frame $X_{0}Y_{0}$, the origin of the reference frame
$X_{0}Y_{0}$ is the instantaneous center of rotation ($ICR$) of
the object.

Substituting (\ref{eq:14}) into (\ref{eq:12}), the differential
equation of centered moments of ellipse when the object is undergoing
coordinated turn motion is given by
\begin{align}
\dot{\eta}_{ij} & =\omega\iint_{R}[\frac{\partial h}{\partial x}(y_{r}-y)+\frac{\partial h}{\partial y}(x-x_{r})]dxdy\nonumber \\
 & =\omega\iint_{R}[(\frac{\partial h}{\partial y}x-\frac{\partial h}{\partial x}y)+(-\frac{\partial h}{\partial y}x_{r}+\frac{\partial h}{\partial x}y_{r})]dxdy\label{eq:dynamic model of moments CT}
\end{align}
The dynamic models of the normalized centered moments of the ellipse
can be calculated using (\ref{eq:dynamic model of moments CT}) as

\begin{equation}
\begin{aligned}\dot{n}_{11}= & \omega(n_{20}-n_{02})\\
\dot{n}_{20}= & -2\omega n_{11}\\
\dot{n}_{02}= & 2\omega n_{11}
\end{aligned}
\end{equation}
The state space representation of the normalized centered moments
of the ellipse $\mathbf{p}_{\mathrm{IM}}=[n_{11},n_{20},n_{02}]^{T}$
are
\begin{equation}
\mathbf{\dot{p}}_{\mathrm{IM}}=\left[\begin{array}{ccc}
0 & \omega & -\omega\\
-2\omega & 0 & 0\\
2\omega & 0 & 0
\end{array}\right]\mathbf{p}_{\mathrm{IM}}\label{eq:state space CT}
\end{equation}
and the solution to the state space in (\ref{eq:state space CT})
is
\begin{equation}
\mathbf{p}_{\mathrm{\mathrm{IM}}}(t)=\mathbf{M}(t,t_{0})\mathbf{p}_{\mathrm{IM}}(t_{0})\label{eq:coordinate turning}
\end{equation}
where the transition matrix $\mathbf{M(t,t_0)} = \left[\begin{smallmatrix} \mathrm{cos}2\theta & \frac{1}{2}\mathrm{sin}2\theta & -\frac{1}{2}\mathrm{sin}2\theta \\ -\mathrm{sin}2\theta & \mathrm{cos}^2\theta & \mathrm{sin}^2\theta \\ \mathrm{sin}2\theta & \mathrm{sin}^2\theta & \mathrm{cos}^2\theta\end{smallmatrix} \right]$,
where $\theta=\omega\left(t-t_{0}\right)$.\textcolor{red}{{} }The derivation
of the transition matrix is shown in the Appendix \ref{sec:appendix A}.

At each time step $k$, the complete state to be tracked is $\mathbf{p}_{k}=[\begin{array}{cc}
\mathbf{p}_{\mathrm{IM},k}^{T}, & \mathbf{p}_{\mathrm{CT},k}^{T}\end{array}]^{T}$, where $\mathbf{p}_{\mathrm{IM},k}$ is a component of the state
corresponding to the image moments, $\mathbf{p}_{\mathrm{CT},k}=[\begin{array}{ccccc}
x_{c,k}, & \dot{x}_{c,k}, & y_{c,k}, & \dot{y}_{c,k}, & \omega_{k}\end{array}]^{T}$ is the vector that includes the position, velocity of the centroid
of the extended object and the turning rate of the extended object.
The state equation is given as follows\vspace{-6pt}

\begin{equation}
\mathbf{p}_{k+1}=\mathbf{F}_{\mathrm{CT}}\mathbf{p}_{k}+\mathbf{\Gamma}\mathbf{w}_{k}\label{eq:dynamic CT-1}
\end{equation}
where the state transition matrix $\mathbf{F}_{\mathrm{CT}}=\mathbf{diag}(\mathbf{M},\mathbf{A}),$
$\mathbf{M} = \left[\begin{smallmatrix} \mathrm{cos}2\omega_kT & \frac{1}{2}\mathrm{sin}2\omega_kT & -\frac{1}{2}\mathrm{sin}2\omega_kT \\ -\mathrm{sin}2\omega_kT & \mathrm{cos}^2\omega_kT & \mathrm{sin}^2\omega_kT \\ \mathrm{sin}2\omega_kT & \mathrm{sin}^2\omega_kT & \mathrm{cos}^2\omega_kT\end{smallmatrix} \right]$
is obtained from (\ref{eq:coordinate turning}), $T$ is a sampling
period, $\mathbf{A} = \left[\begin{smallmatrix} 1 & \frac{\mathrm{sin}\omega_kT}{\omega_k} & 0 & -\frac{1-\mathrm{cos}\omega_kT}{\omega_k}& 0 \\ 0 & \mathrm{cos}\omega_kT & 0 & -\mathrm{sin}\omega_kT & 0 \\ 0 & \frac{1-\mathrm{cos}\omega_kT}{\omega_k} & 1 & \frac{\mathrm{sin}\omega_kT}{\omega_k} & 0 \\ 0 & \mathrm{sin}\omega_kT & 0 & \mathrm{cos}\omega_kT & 0 \\ 0 & 0 & 0  & 0 &1  \end{smallmatrix} \right]$,
$\Gamma=\mathbf{diag}(\begin{array}{cc}
\mathbf{I}_{3\times3}, & \Gamma_{\mathrm{CT}}\end{array})$ with $\Gamma_\mathrm{CT} = \left[\begin{smallmatrix} \frac{1}{2}T^2&0&0 \\ T&0&0 \\ 0&\frac{1}{2}T^2&0 \\ 0&T&0\\0&0&T \end{smallmatrix} \right]$,
and $\mathbf{w}_{k}\in\mathbb{R}^{6\times1}$ is the zero-mean Gaussian
noise vector. Notice that this model is piece-wise continuous.

\subsection{Measurement Model}

Assuming the uniformly generated measurement $\mathbf{\bar{z}}=[\begin{array}{cc}
x, & y\end{array}]^{T}$ without the sensor noise, (\ref{eq:moment_ellipse_cartesian}) maps
the unknown parameters $\mathbf{p}$ to the pseudo-measurement $0$
with the squared scale term $s^{2}\sim\mathcal{U}(0,1)$. The scaling
factor $s$ is approximated to be Gaussian distributed with mean $\nicefrac{2}{3}$
and variance $\nicefrac{1}{18}$ \cite{wahlstrom2015extended}.  Consider
the real measurement $\mathbf{z}=[\begin{array}{cc}
\tilde{x}, & \tilde{y}\end{array}]^{T}$ of the unknown true measurement $\mathbf{\bar{z}}=[\begin{array}{cc}
x, & y\end{array}]^{T}$ in the presence of the additive white Gaussian noise $\boldsymbol{\mathbf{\nu}}=[\nu_{x},\nu_{y}]^{T}$,
where $\nu_{x}\sim\mathcal{N}(\begin{array}{cc}
0, & \sigma_{x}^{2}\end{array})$ and $\nu_{y}\sim\mathcal{N}(\begin{array}{cc}
0, & \sigma_{y}^{2}\end{array})$, the real measurement $\mathbf{z}$ can be expressed as $\mathbf{z}=\mathbf{\bar{z}}+\mathbf{\boldsymbol{\nu}}$.
To find the relationship between the state vector $\mathbf{p}$ and
the real measurement $\mathbf{z}=[\begin{array}{cc}
\tilde{x}, & \tilde{y}\end{array}]^{T}$, the measurement model is derived by substituting $\mathbf{z}$ in
(\ref{eq:moment_ellipse_cartesian}). The following expression can
be obtained\vspace{-4pt}
\begin{align}
g(\bar{\mathbf{z}},\mathbf{p})=g(\mathbf{z},\mathbf{p})-f(\mathbf{z},\mathbf{\boldsymbol{\nu},p})=v\label{eq:ObservationModel}
\end{align}
where $v$ is the pseudo-measurement with the true value of $0$ and
$f(\mathbf{z},\mathbf{\boldsymbol{\nu},p})$ is a polynomial related
to the white noise $\mathbf{\boldsymbol{\nu}}$, which has the mean

\begin{equation}
E\left[f(\mathbf{z},\mathbf{\boldsymbol{\nu},p})\right]=\rho(n_{02}\sigma_{x}^{2}+n_{20}\sigma_{y}^{2})\label{eq:mean}
\end{equation}
and covariance as
\begin{equation}
\begin{aligned}C_{f(\mathbf{z},\mathbf{\boldsymbol{\nu},p})} & =\rho^{2}\biggl\{2n_{02}^{2}\sigma_{x}^{4}+2n_{20}^{2}\sigma_{y}^{4}+4n_{11}^{2}\sigma_{x}^{2}\sigma_{y}^{2}\\
 & +4\left[n_{02}(\widetilde{x}-x_{g})-n_{11}(\widetilde{y}-y_{g})\right]^{2}\sigma_{x}^{2}\\
 & +4\left[n_{20}(\widetilde{y}-y_{g})-n_{11}(\widetilde{x}-x_{g})\right]^{2}\sigma_{y}^{2}\biggr\}
\end{aligned}
\label{eq:covariance}
\end{equation}
where $\rho=\nicefrac{1}{4\left(n_{20}n_{02}-n_{11}^{2}\right)}$.
The derivation of $f(\mathbf{z},\mathbf{\boldsymbol{\nu},p})$ and
its first two moments are shown in the Appendix \ref{sec:appendix B}.
Since the measurement model is highly nonlinear, the UKF presented
in next section, is used to estimate the state vector $\mathbf{p}$. 

\section{UKF for Extended Object Tracking using Image Moments Based RHM\label{sec:Unscented-Kalman-Filter}}

On the basis of the dynamic motion models and the measurement model,
a recursive Bayesian state estimator for tracking the elliptic extended
objects is derived. The state vector of the elliptic extended object
is $\mathbf{p}$. At each time step, several measurement points from
the volume or area of the object's extent are received. The task of
the Bayesian state estimator is to perform backward inference, inferring
the true state parameters from the measurement points. The measurement
points at time step $k$ is denoted as $\mathbf{Z}_{k}=\{\mathbf{z}_{k,l}\}_{l=1}^{L_{k}}$,
assuming there are $L_{k}$ measurements at time $k$ and each measurement
point is $\mathbf{z}_{k,l}=[\begin{array}{cc}
x, & y\end{array}]^{T}$. The state vector up to time step $k$ when all the measurements
are incorporated is denoted as $\mathbf{p}_{k}$. Suppose that the
posterior probability density function (pdf) $p(\mathbf{p}_{k-1}\mid\mathbf{Z}_{k-1})$
at time step $k-1$ is available, the prediction $p(\mathbf{p}_{k}\mid\mathbf{Z}_{k-1})$
for time step $k$ is given by the Chapman-Kolmogorov equation as
\cite{bar2011tracking}

\begin{equation}
p(\mathbf{p}_{k}\mid\mathbf{Z}_{k-1})=\int p(\mathbf{p}_{k}\mid\mathbf{p}_{k-1})p(\mathbf{p}_{k-1}\mid\mathbf{Z}_{k-1})d\mathbf{p}_{k-1}
\end{equation}
the state vector evolves by the conditional density function $p(\mathbf{p}_{k}\mid\mathbf{p}_{k-1})$.
Assuming the Markov model is conformed, the conditional density function
$p(\mathbf{p}_{k}\mid\mathbf{p}_{k-1})$ can be derived based on different
dynamic models in Subsection \ref{subsec:Dynamic-model}. Assuming
the measurements $\mathbf{Z}_{k}=\{\mathbf{z}_{k,l}\}_{l=1}^{L_{k}}$
at time $k$ are independent, the prediction $p(\mathbf{p}_{k}\mid\mathbf{z}_{k,l})$
is updated recursively via Bayes rule as

\begin{equation}
p(\mathbf{p}_{k}\mid\mathbf{z}_{k,l})\wasypropto p(\mathbf{z}_{k,l}\mid\mathbf{p}_{k})p(\mathbf{p}_{k}\mid\mathbf{z}_{k,l-1})
\end{equation}
where $p(\mathbf{p}_{k}\mid\mathbf{z}_{k,0})=p(\mathbf{p}_{k}\mid\mathbf{Z}_{k-1})$
and $p(\mathbf{p}_{k}\mid\mathbf{Z}_{k})=p(\mathbf{p}_{k}\mid\mathbf{z}_{k,L_{k}})$. 

When the target is moving with uniform motion (constant velocity model,
which is a linear system), its states $\mathbf{p}_{k|k-1}$ and covariance
$C_{k|k-1}$ are predicted based on the dynamic model (\ref{eq:cvm-1})
as\vspace{-4pt}
\begin{align}
\mathbf{p}_{k|k-1} & =\mathbf{F}_{\mathrm{CV}}\mathbf{p}_{k}\label{eq:predict_state}\\
C_{k|k-1} & =\mathbf{F}_{\mathrm{CV}}\mathbf{p}_{k}\mathbf{F}_{\mathrm{CV}}^{T}+\mathbf{C}_{\mathrm{CV}}\label{eq:predict_covariance}
\end{align}
However, the proposed image moments based RHM and its dynamic model
such as coordinated turn model are nonlinear. When the system is nonlinear,
the linearization method like the extended Kalman filter (EKF) will
introduce large errors in the true posterior mean and covariance.
UKF addresses this problem by the method of unscented transformation
(UT), which doesn't require the calculations of the Jacobian and Hessian
matrices. The UT sigma point selection scheme results in approximations
that are accurate to the third order for Gaussian inputs for all nonlinearities
and has the same order of the overall number of computations as the
EKF \cite{wan2000unscented}. When the state variables in $\mathbf{p}\in\mathbb{R}^{M\times1}$
with mean $\mathbf{\bar{p}}$ and covariance $C_{\mathbf{p}}$ are
propagating through a nonlinear function $\mathbf{y}=f(\mathbf{p})$,
such as (\ref{eq:coordinate turning}) or (\ref{eq:ObservationModel}),
the mean $\bar{\mathbf{y}}$ and covariance $C_{\mathbf{y}}$ of $\mathbf{y}$
are approximated by generating the UT sigma points $\mathcal{X}_{i}$
as \cite{wan2000unscented}

\vspace{-10 pt}

\begin{align}
\bar{\mathbf{y}} & =\sum_{i=0}^{2M}W_{i}^{(m)}\mathcal{Y}_{i}\label{eq:28mean}\\
C_{\mathbf{y}} & =\sum_{i=0}^{2M}W_{i}^{(C)}\left\{ \mathcal{Y}_{i}-\bar{\mathbf{y}}\right\} \left\{ \mathcal{Y}_{i}-\bar{\mathbf{y}}\right\} \label{eq:mean-covariance-sigma}
\end{align}
where $\mathcal{Y}_{i}=f(\mathbf{\mathcal{X}}_{i})$. The sigma points
$\mathcal{X}_{i}$ and the weights $W_{i}^{(m)}$ and $W_{i}^{(C)}$
are calculated by \cite{wan2000unscented}

\vspace{-10 pt}

\begin{equation}
\begin{aligned}\mathcal{X}_{0} & =\mathbf{\bar{p}}\\
\mathcal{X}_{i} & =\mathbf{\bar{p}}+\left(\sqrt{(M+\lambda)C_{\mathbf{p}}}\right)_{i} & i=1,\ldots,M\\
\mathcal{X}_{i} & =\mathbf{\bar{p}}-\left(\sqrt{(M+\lambda)C_{\mathbf{p}}}\right)_{i} & i=M+1,\ldots2M\\
W_{0}^{(m)} & =\lambda/(M+\lambda)\\
W_{0}^{(C)} & =\lambda/(M+\lambda)+(1-\alpha^{2}+\beta)\\
W_{i}^{(m)} & =W_{i}^{(C)}=1/\left[2(M+\lambda)\right] & i=1,\ldots,2M
\end{aligned}
\label{eq:sigma points}
\end{equation}
where $\lambda$ is the scaling parameter as $\lambda=\alpha^{2}(M+\kappa)-M$,
$\alpha$ is the parameter determines the spread of the sigma points
around the mean $\bar{\mathbf{p}}$, $\kappa$ is the secondary scaling
parameter usually set to $0$ and $\beta$ is the parameter to incorporate
the prior knowledge of the distribution of $\mathbf{p}$. The UKF
for image moments based random hypersurface model is illustrated in
Algorithm. \ref{alg:UKF-with-sequential}.

\begin{algorithm}[h]
Set the time steps ${N}$\; 

Set the initial state vector $\mathbf{p}_{0}$ and covariance $C_{0}$
\; 

\For{$k$=1 to N}{

\Case{Constant velocity model}{

State $\mathbf{p}_{k|k-1}$ is predicted as in (\ref{eq:predict_state});

Covariance $C_{k|k-1}$ is predicted as in (\ref{eq:predict_covariance});

}

\Case{Coordinated turn model}{

Augment the state vector $\mathbf{p}_{k-1}^{a}=\left[\left(\mathbf{p}_{k-1}\right)^{T},\mathbf{w}_{k}^{T}\right]^{T}$;

Calculate sigma points using (\ref{eq:sigma points});

States prediction based on (\ref{eq:dynamic CT-1}) with sigma points;

Using (\ref{eq:28mean}), (\ref{eq:mean-covariance-sigma}) to calculate
the mean and covariance of the state vector $\mathbf{p}_{k|k-1}$;

}

Obtain the measurement points $\mathbf{Z}_{k}=\{\mathbf{z}_{k,l}\}_{l=1}^{L_{k}}$
at time step $k$\; 

\For{$l$=1 to $L_{k}$}{

Calculate the mean and covariance of $f(\mathbf{z}_{k,l},\mathrm{\mathrm{\mathbf{\boldsymbol{\nu}}}},\mathbf{p}_{k|k-1})$
using (\ref{eq:mean}) and (\ref{eq:covariance})\; 

Augment the state vector $\mathbf{p}_{k|k-1,l}^{a}=\left[\left(\mathbf{p}_{k|k-1,l}\right)^{T},f(\mathbf{z}_{k,l},\mathrm{\mathrm{\mathbf{\boldsymbol{\nu}}}},\mathbf{p}_{k|k-1}),s\right]^{T}$\; 

Calculate sigma points using (\ref{eq:sigma points})\; 

Pseudo-measurement $v_{k,l}$ calculated based on (\ref{eq:ObservationModel})
for measurement point $\mathbf{z}_{k,l}$\; 

Using (\ref{eq:28mean}), (\ref{eq:mean-covariance-sigma}) to calculate
the mean and covariance of the $v_{k,l}$\; 

Update state vector $\mathbf{p}_{k,l}$\; 

}

}

\caption{UKF with sequential processing of measurements\label{alg:UKF-with-sequential}.}
\end{algorithm}

\section{Tracking extended target with IMM\label{sec:Tracking-extended-target with IMM}}

The proposed image moments based random hypersurface model is embedded
with the IMM approach for tracking extended target undergoing complex
trajectories in this section. When the extended target is switching
between maneuvering and non-maneuvering behaviors, its kinematic state
and spatial extent may change abruptly. Multiple model approaches,
such as interacting multiple model (IMM), are effective to track the
target with complex trajectories, especially with high maneuvering
index (larger than $0.5$) \cite{kirubarajan2003kalman,bar2004estimation,granstrom2015systematic}.
The IMM approach assumes the target obeys one of a finite number of
motion models and identifies the beginning and the end of the motion
models by updating the model probabilities. The adaptation via model
probability update helps the IMM approach keep the estimation errors
consistently low, both during maneuvers as well as no-maneuver intervals.
Details about the IMM for point target tracking can be found in literature
such as \cite{bar2004estimation}. 

\begin{figure}
\centering{}\includegraphics[width=0.8\columnwidth]{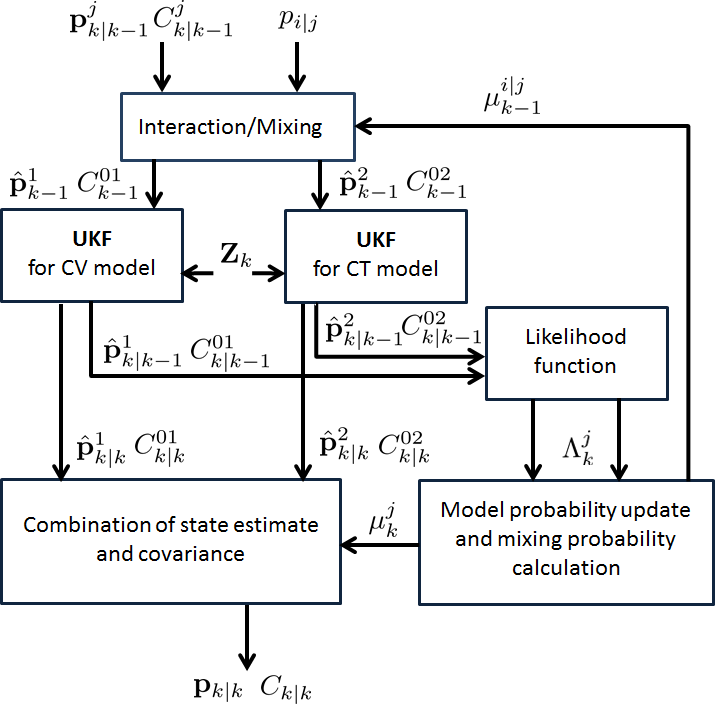}\caption{Flowchart of UKF-IMM framework.\label{fig:Flowchart-of-UKF-IMM}}
\end{figure}

The proposed image moments based random hypersurface model with the
dynamic motion models, such as the constant velocity motion model
and the coordinated turn motion model in Section \ref{sec:Image-Moment-based-RHM},
are integrated in an IMM framework. Since the dynamic motion model
and the measurement model are nonlinear, the UKF-IMM algorithm is
proposed. The flowchart of the UKF-IMM algorithm are shown in Fig.
\ref{fig:Flowchart-of-UKF-IMM}, where $\mu_{k-1}^{i|j}$ is the mixing
probability, $p_{i|j}$ is the Markov chain transition matrix between
the $i\mathrm{th}$ and $j\mathrm{th}$ models and $\Lambda_{k}^{j}$
are likelihood function corresponding to the $j\mathrm{th}$ model.
There are multiple measurement points at each time step, the sequential
approach is adopted for UKF and the likelihood function is generated
based on the measurement model. 

At each time step, assuming there are $L_{k}$ measurements $\mathbf{Z}_{k}=\{\mathbf{z}_{k,l}\}_{l=1}^{L_{k}}$.
The pseudo-measurement variable $v_{k,l}$ can be generated for each
measurement $\mathbf{z}_{k,l}$, based on the predicted state vector
$\mathbf{p}_{k|k-1}^{j}$, covariance $C_{k|k-1}^{j}$ and the measurement
model in (\ref{eq:ObservationModel}). The mean and the covariance
of the pseudo-measurement variable $v_{k,l}$, can be obtained by
the method of unscented transformation (UT). Assuming the measurements
$\mathbf{Z}_{k}=\{\mathbf{z}_{k,l}\}_{l=1}^{L_{k}}$ are independent
identically Gaussian distributed, the log-likelihood function based
on the pseudo-measurement variable $v_{k,l}$ is 
\begin{equation}
\log\Lambda_{k}^{j}=\sum_{l=1}^{L_{k}}\left[-\frac{(0-\mu_{v,l})^{2}}{2\sigma_{v,l}^{2}}-\log\left(\sqrt{2\pi\sigma_{v,l}^{2}}\right)\right]\label{eq:log-likelihood}
\end{equation}
where $\mu_{v,l}$ and $\sigma_{v,l}^{2}$ are the mean and covariance
of the pseudo-measurement $v_{k,l}$, generated for each measurement
point $\mathbf{z}_{k,l}$. In many cases, the likelihood $\Lambda^{j}\left(k\right)$
can become extremely small. To avoid this issue, the average log-likelihood
$\log\bar{\Lambda}^{j}\left(k\right)$ is used which is given by
\begin{equation}
\log\bar{\Lambda}_{k}^{j}=\frac{1}{n_{k}}\log\Lambda_{k}^{j}\label{eq:AV log-likelihood 1}
\end{equation}
and 
\begin{equation}
\bar{\Lambda}_{k}^{j}=\exp\left(\log\bar{\Lambda}_{k}^{j}\right)\label{eq:AV log-likelihood 2}
\end{equation}
which is the value of the measurement likelihood between $0$ and
$1$. This measurement likelihood is used in the IMM filter. The details
of the calculation of the measurement likelihood is show in Algorithm
\ref{alg:likelihood function}.

\begin{algorithm}
Obtain the predicted state vector $\mathbf{p}_{k|k-1}^{j}$ and covariance
$C_{k|k-1}^{j}$ of model $j$\; 

Obtain the measurement points $\mathbf{Z}_{k}=\{\mathbf{z}_{k,l}\}_{l=1}^{L_{k}}$
at time step $k$\; 

\For{$l$=1 to $L_{k}$}{

Calculate the mean and covariance of $f(\mathbf{z}_{k,l},\mathrm{\mathrm{\mathbf{\boldsymbol{\nu}}}},\mathbf{p}_{k|k-1})$
using (\ref{eq:mean}) and (\ref{eq:covariance})\; 

Augment the state vector $\mathbf{p}_{k|k-1,l}^{a}=\left[\left(\mathbf{p}_{k|k-1,l}^{j}\right)^{T},f(\mathbf{z}_{k,l},\mathrm{\mathbf{\boldsymbol{\nu}}},\mathbf{p}_{k|k-1}),s\right]^{T}$\; 

Calculate sigma points $\mathcal{X}$ using (\ref{eq:sigma points})\; 

Propagate sigma points $\mathcal{X}$ through the measurement model
in (\ref{eq:ObservationModel})\; 

Using (\ref{eq:28mean}), (\ref{eq:mean-covariance-sigma}) to calculate
the mean and covariance of the pseudo-measurement $v_{k,l}$\; 

Summation of the value of the log-likelihood function using (\ref{eq:log-likelihood})\; 

} 

Calculation of the value of the average log-likelihood function using
(\ref{eq:AV log-likelihood 1}) and the measurement likelihood using
(\ref{eq:AV log-likelihood 2})\; 

\caption{\label{alg:likelihood function}Calculation of the measurement likelihood
$\bar{\Lambda}_{k}^{j}$ corresponding to the $j\mathrm{th}$ model
by unscented transformation.}
\end{algorithm}

\section{Simulation Results\label{sec:Evaluation}}

In this section, several simulation tests are conducted to evaluate
the performance of the proposed image moments based extended object
tracking. To validate the measurement model in (\ref{eq:ObservationModel}),
the shapes of the static objects are estimated with different noise
levels in the first simulation. Then the tracking of the extended
target moving with linear motion and coordinated turn motion are demonstrated.
The constant velocity model in (\ref{eq:cvm-1}) and the nearly coordinated
turn model in (\ref{eq:dynamic CT-1}) are used and validated for
these cases. Two targets with the shapes of the plus-sign and ellipse
are used in the simulations. At last, tracking of targets moving with
maneuvering and non-maneuvering intervals are presented. Two scenarios
are simulated in this test. One with slow motion and maneuvers and
the other with fast motion and maneuvers. The UKF-IMM algorithm with
constant velocity model and the nearly coordinated turn model is applied
in these cases. The RMM and its combination with the IMM in \cite{feldmann2011tracking}
are implemented as a benchmark comparison for our proposed image moments
based random hypersurface model. 

The intersection over union (IoU) is used as the metric to evaluate
the proposed algorithm. The IoU is defined as the area of the intersection
of the estimated shape and the true shape divided by the union of
the two shapes\cite{yang2016metrics}
\begin{equation}
\mathbf{IoU}=\frac{area(\mathbf{p})\cap area(\mathbf{\hat{p}})}{area(\mathbf{p})\cup area(\mathbf{\hat{p}})}
\end{equation}
where $\mathbf{p}$ is the true state vector and $\mathbf{\hat{p}}$
is the estimated state vector. $\mathbf{IoU}$ is between $0$ and
$1$, where the value $1$ corresponds to a perfect match between
the estimated area and the ground-truth. Additionally, the root mean
squared errors (RMSE) of the estimated position and velocity of the
centroid $(x_{c,}y_{c})$ of the extended target are also evaluated,
which are defined as
\begin{equation}
\mathbf{RMSE}=\sqrt{\frac{1}{N}\sum_{i=1}^{N}\xi_{i}^{2}}
\end{equation}
where $N$ is the Monte Carlo runs, $\xi_{i}$ is the error of the
estimation from the $i$th run. For the RMSE of the position, $\xi_{i,p}\triangleq(\hat{x}_{c}-x_{c})^{2}+(\hat{y}_{c}-y_{c})^{2}$,
where $(\hat{x}_{c,}\hat{y}_{c})$ is the estimated centroid of the
extended target and $(x_{c,}y_{c})$ is the ground-truth. Similarly,
for the RMSE of the velocity, the estimation error is defined as $\xi_{i,v}\triangleq(\hat{\dot{x}}_{c}-\dot{x}_{c})^{2}+(\hat{\dot{y}}_{c}-\dot{y}_{c})^{2}$,
where $(\hat{\dot{x}}_{c,}\hat{\dot{y}}_{c})$ is the estimated velocity
of the centroid and $(\dot{x}_{c,}\dot{y}_{c})$ is the ground-truth.

\subsection{Static Extended Objects\label{subsec:Static-Extended-Objects}}

The plus-sign shaped target is made up of two rectangles with the
width and height of $3\mathrm{cm}$ and $0.5\mathrm{cm}$, and $0.5\mathrm{cm}$
and $2\mathrm{cm}$, respectively. The major and minor axes of the
elliptic target are set to $3\mathrm{cm}$ and $2\mathrm{cm}$, respectively.
The simulation is performed by uniformly sampling $400$ points from
the static extended objects. Three different levels of additive Gaussian
white noises with variances such as $\mathbf{diag}(\begin{array}{cc}
0.1^{2}, & 0.1^{2}\end{array})$ (low), $\mathbf{diag}(\begin{array}{cc}
0.5^{2}, & 0.5^{2}\end{array})$ (medium) and $\mathbf{diag}(\begin{array}{cc}
1, & 1\end{array})$ (high) are used to generate the noisy measurements.

\begin{figure*}
\begin{centering}
\makebox[0.8\linewidth][c]{\subfigure[]{\includegraphics[width=0.2\paperwidth]{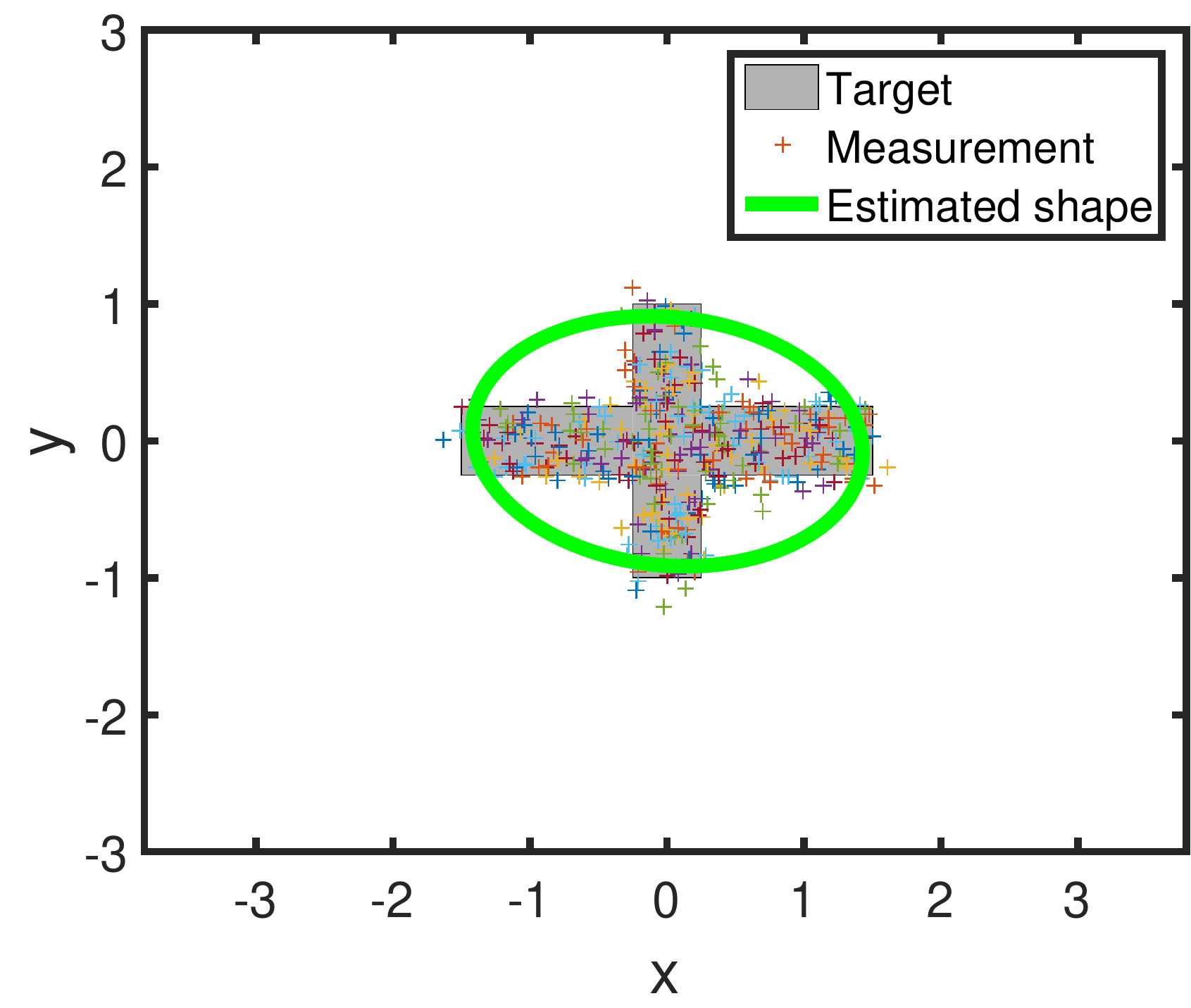}}\hspace{0.4em}\subfigure[]{\includegraphics[width=0.2\paperwidth]{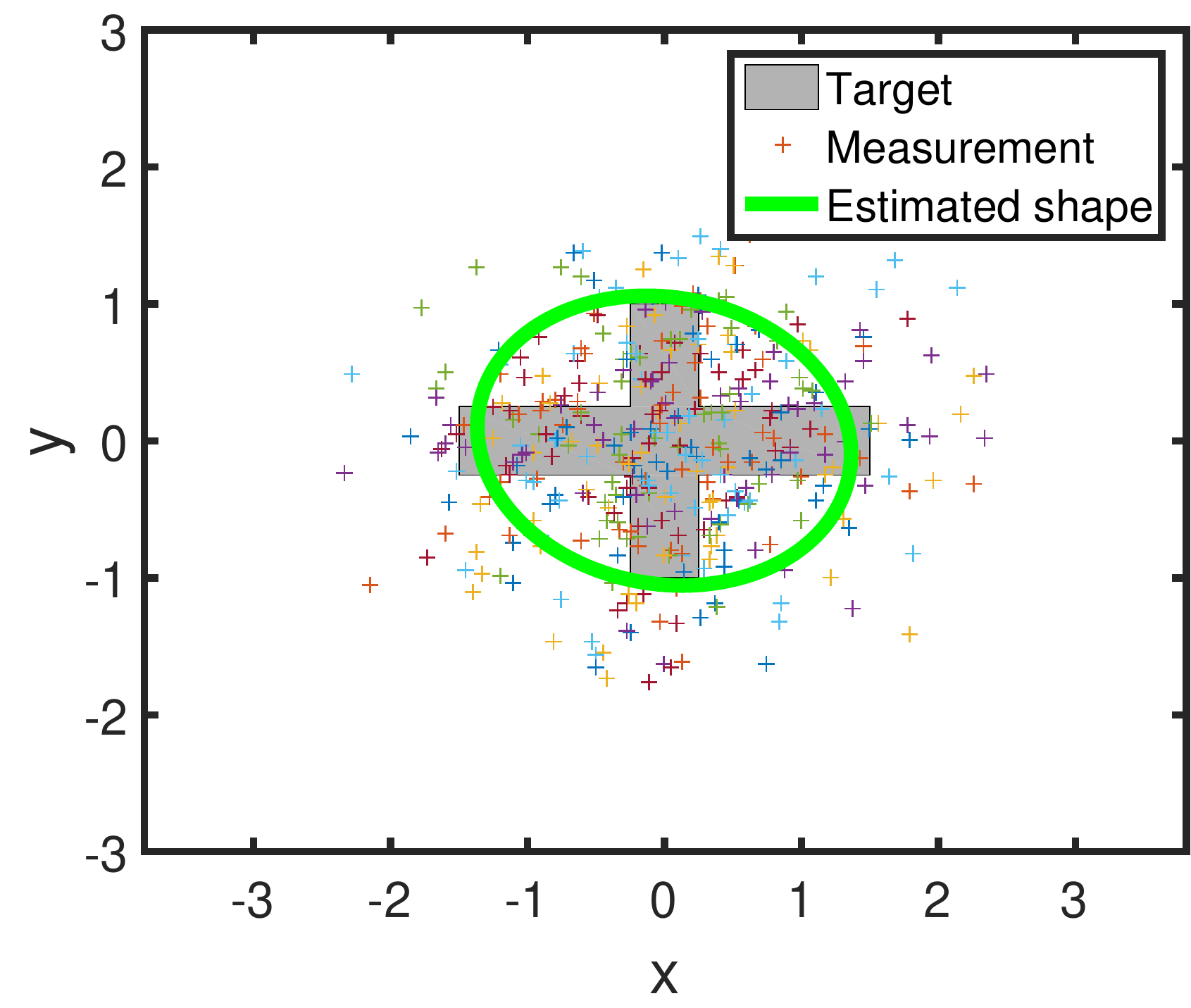}}\hspace{0.4em}\subfigure[]{\includegraphics[width=0.2\paperwidth]{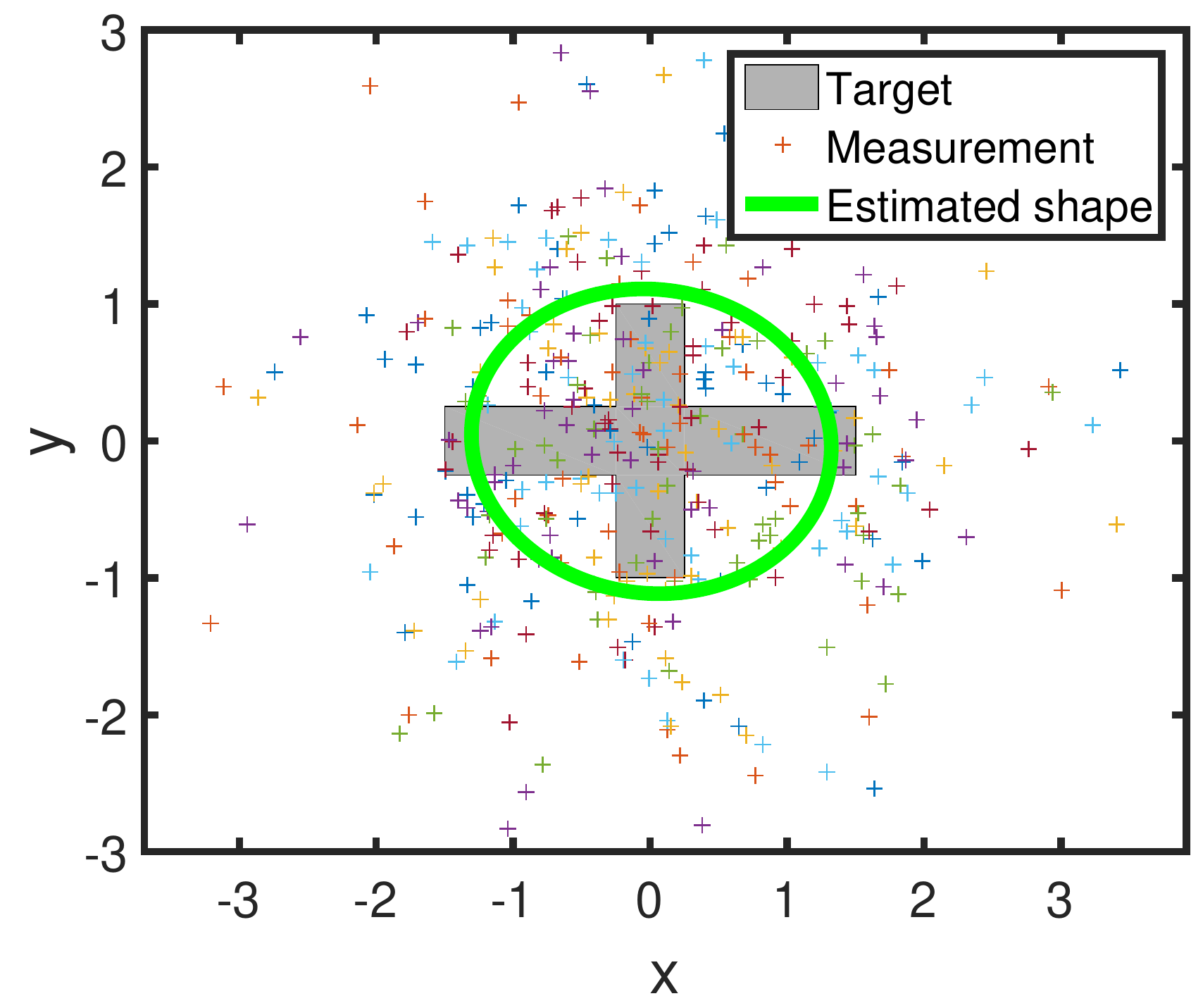}}}\\\vspace{-1.5ex}
\par\end{centering}
\begin{centering}
\makebox[0.8\linewidth][c]{\subfigure[]{\includegraphics[width=0.2\paperwidth]{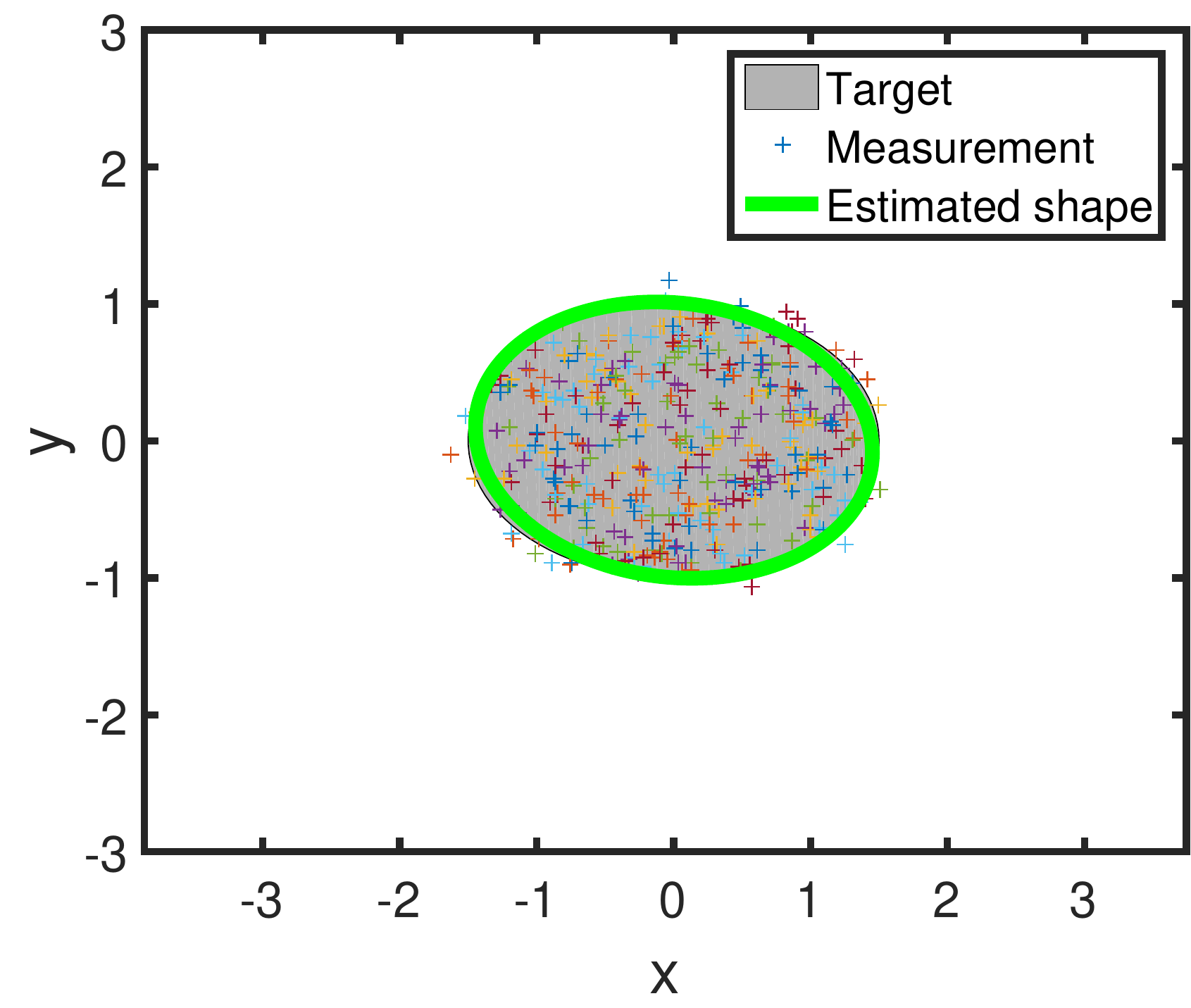}}\hspace{0.4em}\subfigure[]{\includegraphics[width=0.2\paperwidth]{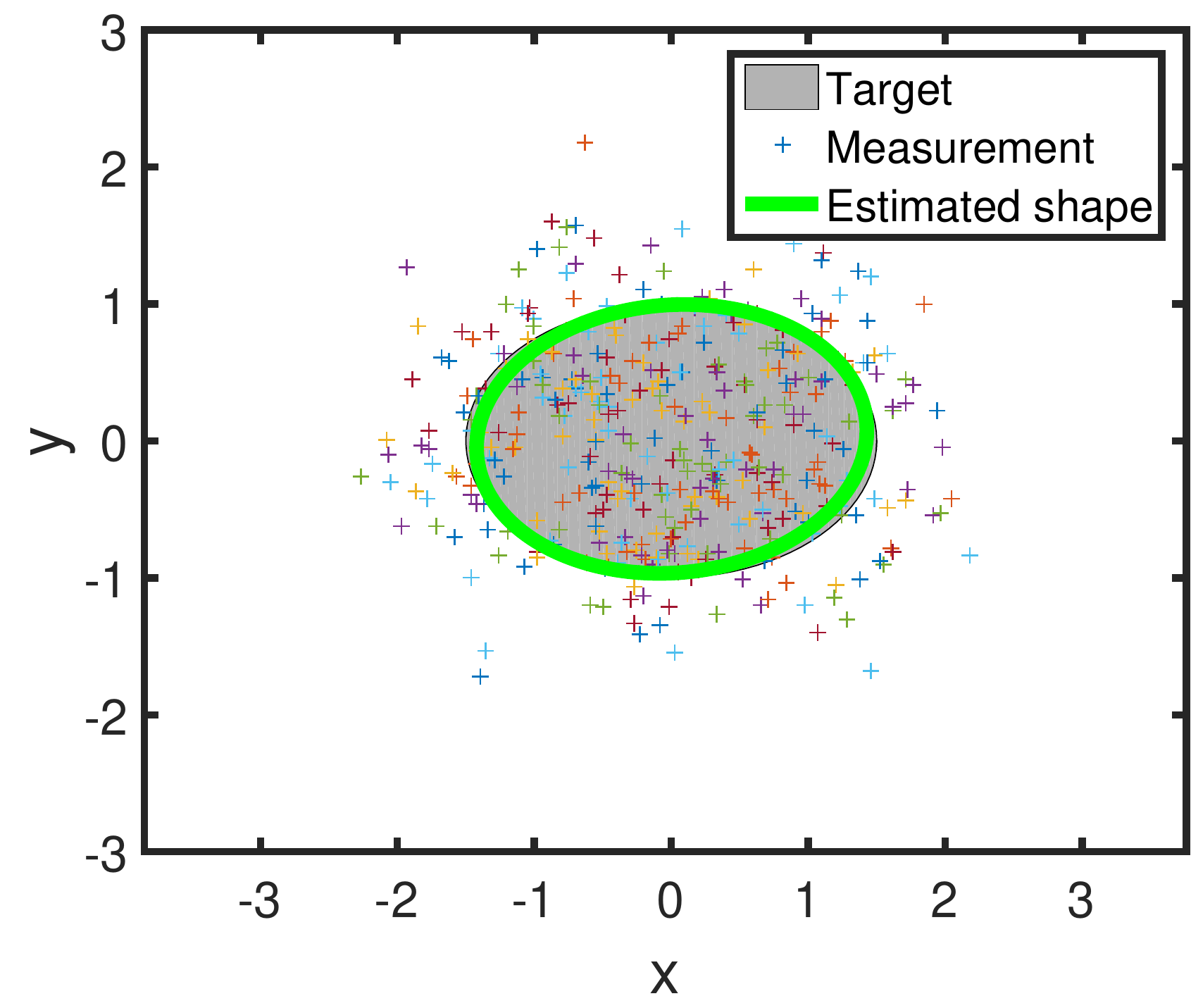}}\hspace{0.4em}\subfigure[]{\includegraphics[width=0.2\paperwidth]{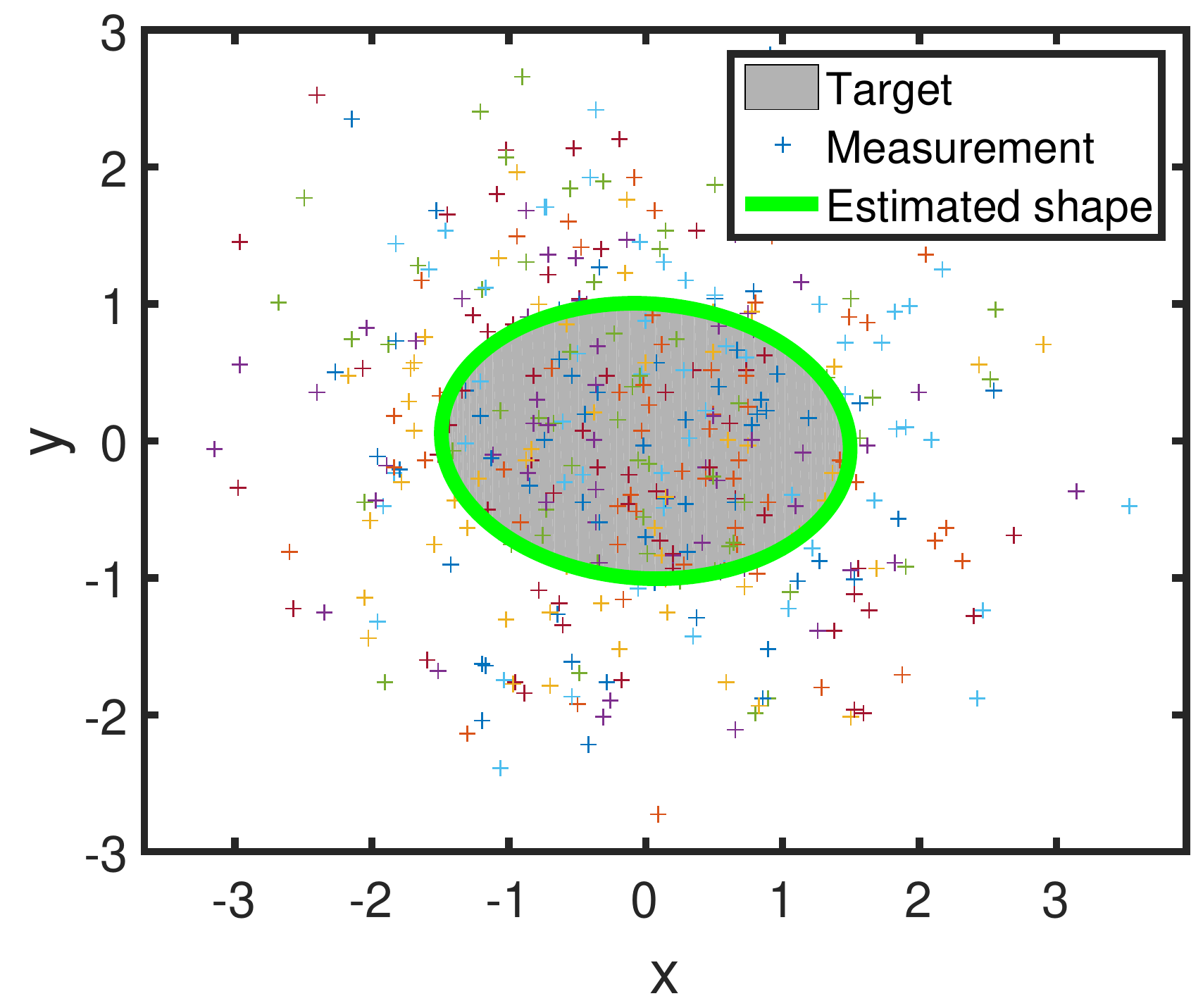}}}\\
\par\end{centering}
\centering{}\caption{Estimation of the shape of extended targets with different measurement
noise levels in a particular run; estimation of the shape is drawn
in green color and the gray shape is the ground truth. (a) - (c) estimations
of the shape of the extended target (with the shape of the plus-sign)
with three different measurement noise levels; (d) - (f) estimations
of the elliptic target with three different measurement noise levels.\label{fig:static}}
\end{figure*}

UFK is used for estimating the state given noisy measurements of points
uniformly sampled from the plus-sign-shaped and ellipse-shaped extended
objects. The state is initialized as a circle with radius of $0.89\mathrm{cm}$
located at the origin. The estimation results for the plus-sign-shaped
object are shown in Figs. \ref{fig:static}(a), \ref{fig:static}(b),
\ref{fig:static}(c) and the estimation results for the ellipse-shaped
object are shown in Figs. \ref{fig:static}(d), \ref{fig:static}(e),
\ref{fig:static}(f). \textcolor{black}{The mean values of the IoU
of the static ellipse and the plus-sign-shaped targets with $3$ different
noise levels are shown in Table}\textcolor{red}{{} }\textcolor{black}{\ref{tab:The-mean-value}.
}The image moments based measurement model can precisely estimate
the shape of the targets. With the increases in covariance of the
measurement noise, the proposed image moments based model also gives
a shape close to the actual shape of the targets. The IoU value for
the plus-sign shaped target is lower than the elliptical target because
the ellipse is used to roughly estimate the plus-sign shape.

\begin{table}[h]
\centering{}\textcolor{black}{\caption{The mean value of\textcolor{black}{{} }Intersection-Over-Union (IoU)
between the true and the estimated target region in different simulated
scenarios. Three noise levels (low , medium and high) are used to
evaluate the static targets (ellipse and plus-sign-shaped target).
The mean value is calculated over 100 Monte Carlo runs.\label{tab:The-mean-value}}
}\resizebox{\columnwidth}{!}{\textcolor{red}{}%
\begin{tabular}{cccc>{\centering}p{0.15\textwidth}>{\centering}p{0.15\textwidth}}
\hline 
\multirow{2}{*}{\textcolor{black}{\large{}Target shape}} & \multicolumn{3}{c}{\textcolor{black}{\large{}Static target}} & \multirow{2}{0.15\textwidth}{\textcolor{black}{\large{}Linear motion}} & \multirow{2}{0.15\textwidth}{\textcolor{black}{\large{}Coordinated turn motion }}\tabularnewline
\cline{2-4} 
 & \textcolor{black}{\large{}Low } & \textcolor{black}{\large{}Medium } & \textcolor{black}{\large{}High } &  & \tabularnewline
\hline 
\hline 
\textcolor{black}{\large{}Ellipse} & \textcolor{black}{\large{}$0.90$} & \textcolor{black}{\large{}0.88} & \textcolor{black}{\large{}0.85} & \textcolor{black}{\large{}0.88} & \textcolor{black}{\large{}0.87}\tabularnewline
\hline 
\textcolor{black}{\large{}Plus-sign} & \textcolor{black}{\large{}$0.48$} & \textcolor{black}{\large{}0.48} & \textcolor{black}{\large{}0.47} & \textcolor{black}{\large{}0.46} & \textcolor{black}{\large{}0.48}\tabularnewline
\hline 
\end{tabular}} 
\end{table}

\begin{figure*}
\begin{centering}
\makebox[0.8\linewidth][c]{\subfigure[ellipse]{\includegraphics[width=0.17\paperwidth]{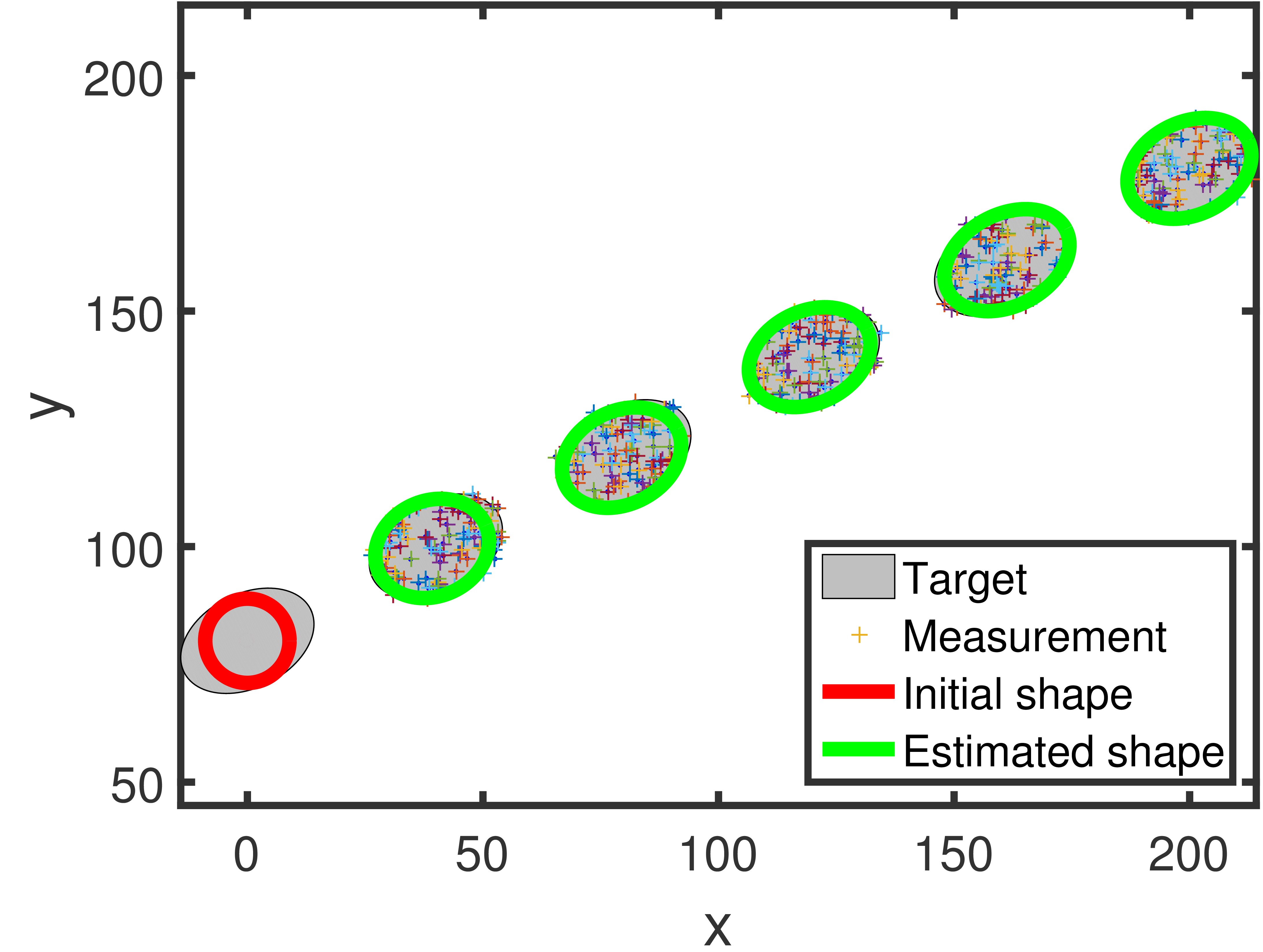}}\hspace{1em}\subfigure[plus-sign]{\includegraphics[width=0.17\paperwidth]{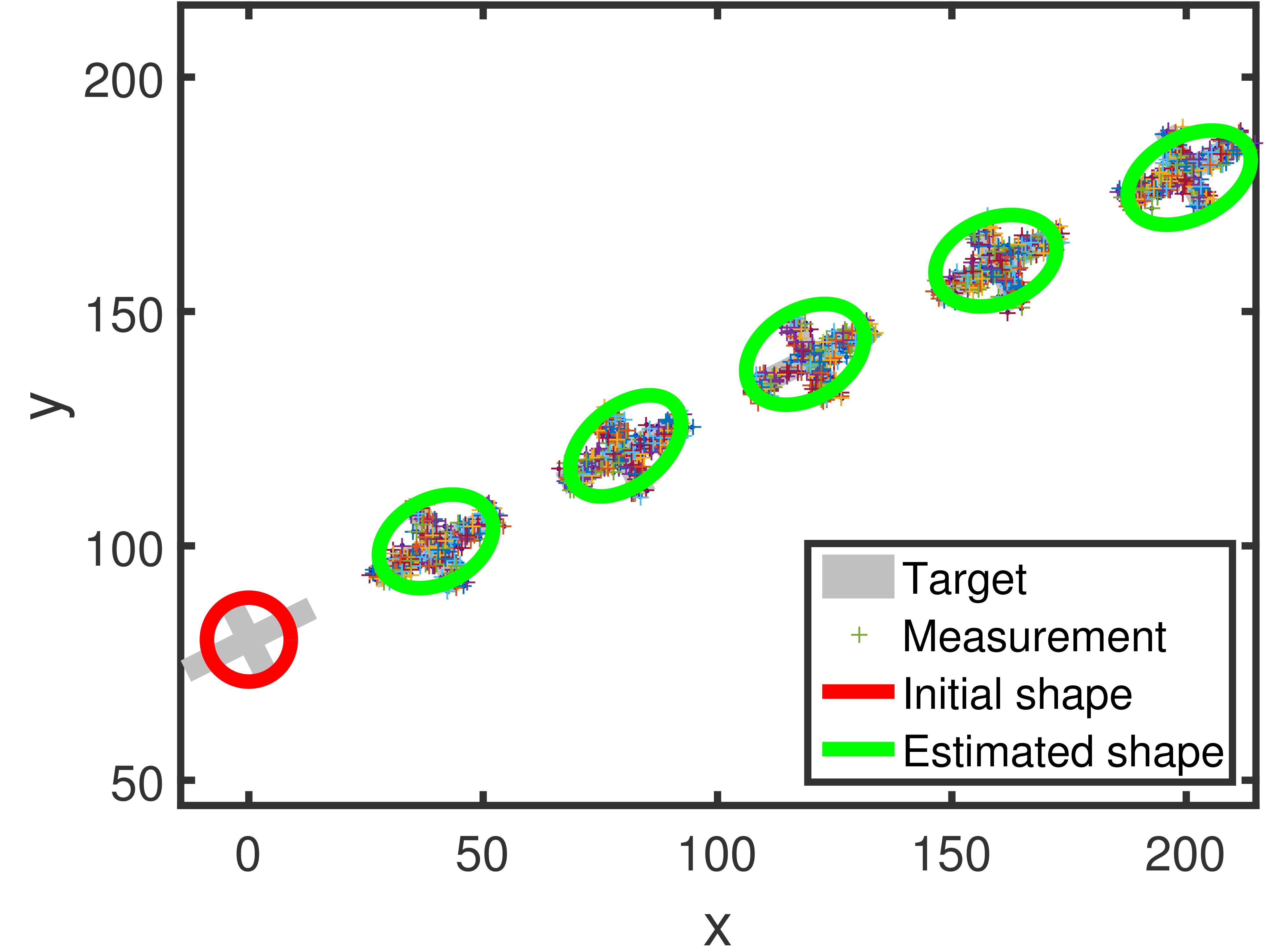}}\hspace{1em}\subfigure[ellipse]{\includegraphics[width=0.17\paperwidth]{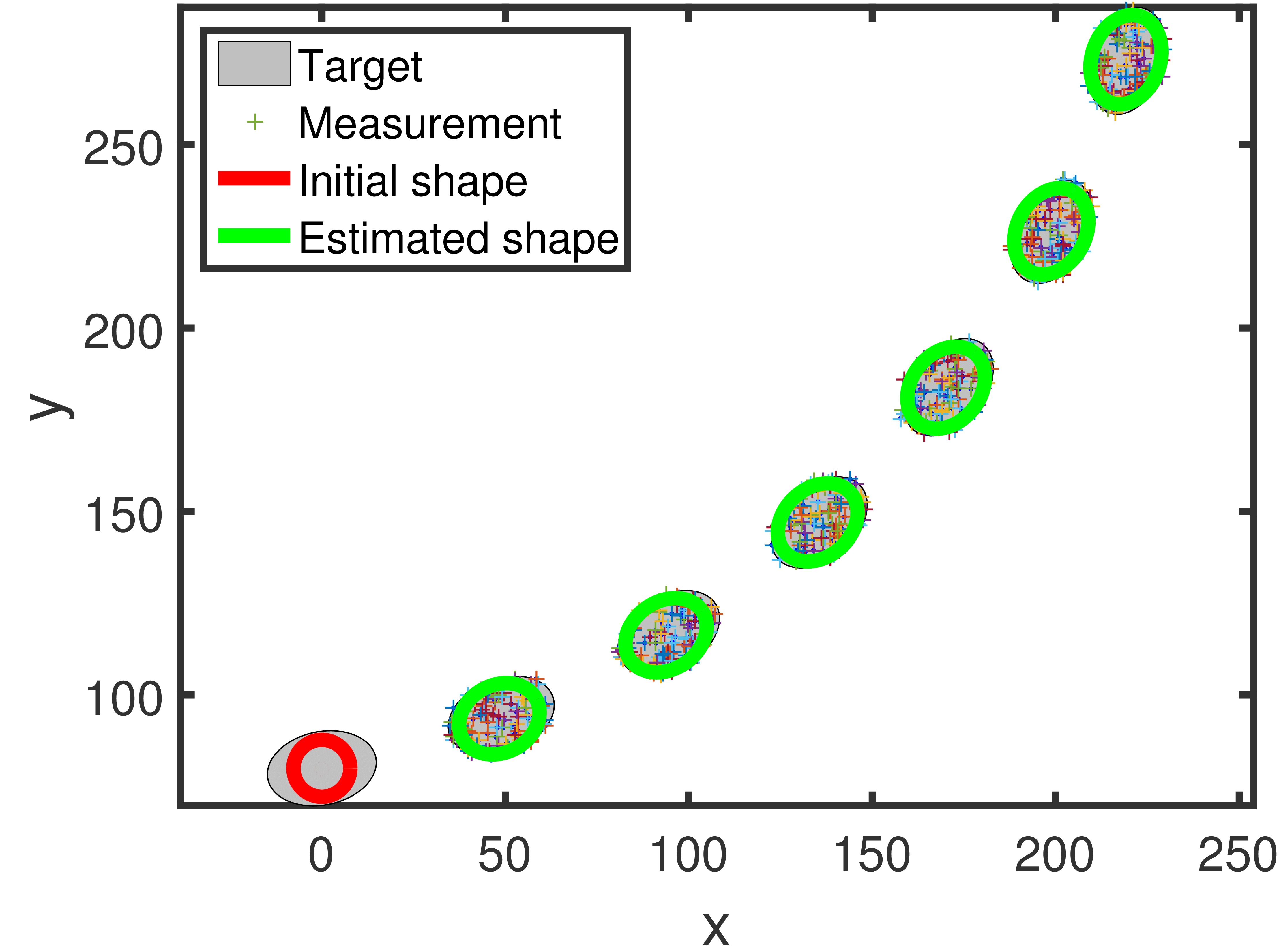}}\hspace{1em}\subfigure[plus-sign]{\includegraphics[width=0.17\paperwidth]{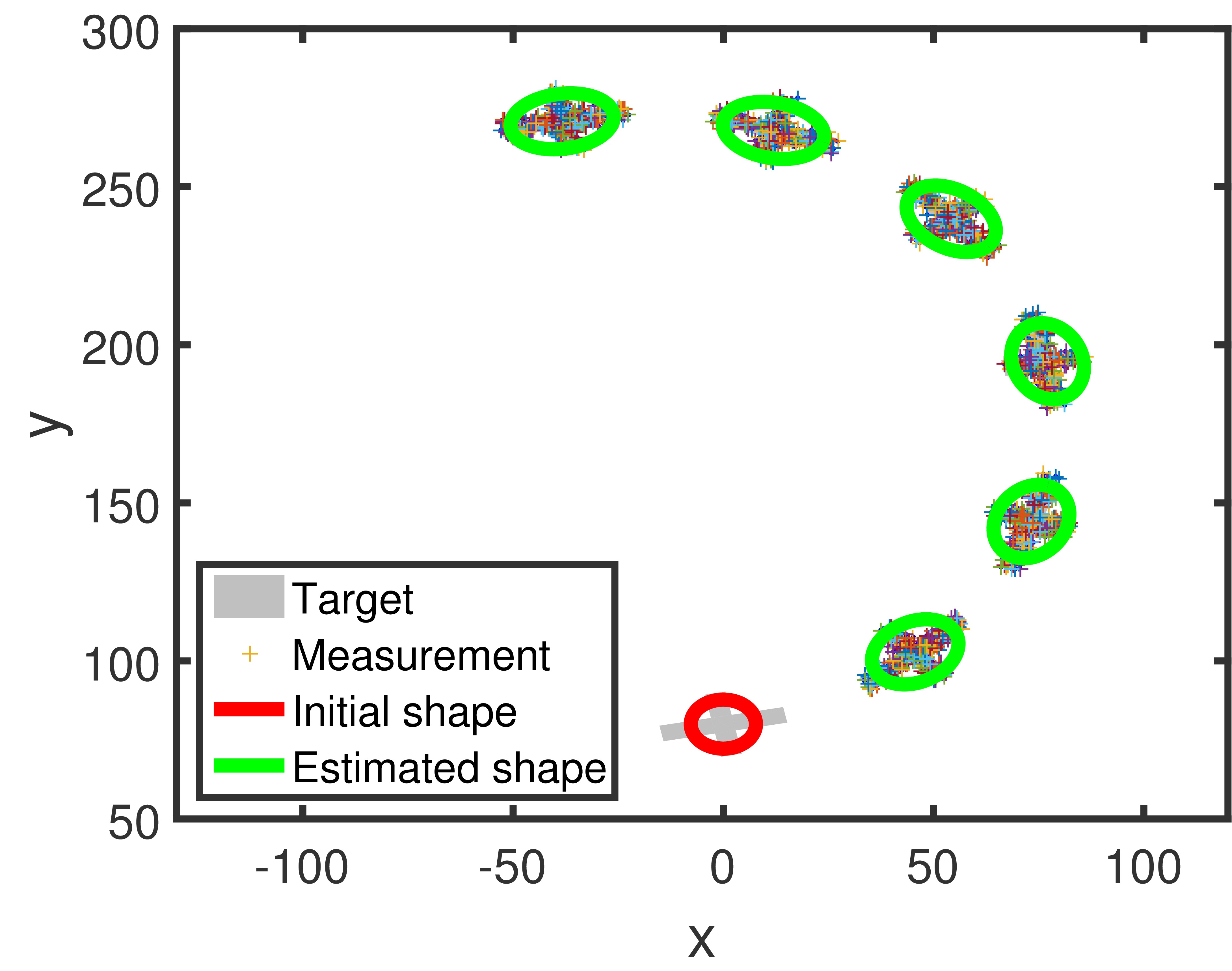}}}\\
\par\end{centering}
\centering{}\caption{Tracking of the extended objects during constant velocity model or
coordinated turn model in a particular run; Initial shape is shown
in red color, estimation of the shape is drawn in green color and
the gray shape is the ground truth; (a) ellipse with a constant velocity;
(b) plus-sign with a constant velocity; (c) the ellipse executes a
$1^\circ/s$ coordinated turn; (d) the plus-sign executes a $3^\circ/s$
coordinated turn.\label{fig:constant velocity model} }
\end{figure*}

\subsection{Linear Motion \label{subsec:Constant-velocity-model}}

In this subsection, extended objects with plus-sign and elliptical
shapes moving with a constant velocity are simulated. The plus-sign
shaped target is made up of two rectangles with the width and height
of $30\mathrm{cm}$ and $5\mathrm{cm}$, and $5\mathrm{cm}$ and $20\mathrm{cm}$,
respectively. For the ellipse-shaped object, the major and minor axes
are set to $30\mathrm{cm}$ and $20\mathrm{cm}$, respectively. The
extended objects start moving from position $[\begin{array}{cc}
0, & 80\end{array}]^{T}$$\mathrm{cm}$ with a constant velocity of $[\begin{array}{cc}
4, & 2\end{array}]^{T}$$\unitfrac{cm}{s}$ for $60$ seconds and the measurements are generated
from the targets at every $10$ seconds. At each time step $k$, $100$
measurement points uniformly sampled from the objects are generated. 

For UKF implementation, the states are initialized as a circle with
radius of $8.9\mathrm{cm}$, located at $[\begin{array}{cc}
0, & 80\end{array}]^{T}$$\mathrm{cm}$ with a constant velocity of $[\begin{array}{cc}
4, & 2\end{array}]^{T}$$\unitfrac{cm}{s}$. The Gaussian white noise variance is selected
as $\mathbf{diag}(\begin{array}{cc}
1 & 1\end{array})$ for each point measurement. The parameter $q$ for the process noise
covariance in the constant velocity model in (\ref{eq:cvm-1}) is
set as $q=0.2$ and $\mathbf{C}_{\mathrm{IM}}=\mathrm{\mathbf{diag}}(0.1,0.1,0.1)$.\textbf{
}The tracking results for ellipse-shaped extended object are shown
in Fig. \ref{fig:constant velocity model}(a) and for plus-sign-shaped
extended object are shown in Fig. \ref{fig:constant velocity model}(b).
It can be seen that the shapes of targets are being estimated accurately
as more measurements are obtained. \textcolor{black}{The mean value
of the RMSE of the position over $100$ Monte Carlo runs is $0.58\mathrm{cm}$
for ellipse and $1.29\mathrm{cm}$ for the plus-sign. The mean value
of the RMSE of the velocity over $100$ Monte Carlo runs is $0.50\unitfrac{cm}{s}$
for ellipse and $0.25\unitfrac{cm}{s}$ for the plus-sign. The mean
values of the IoU of the ellipse and the plus-sign-shaped targets
during the linear motion are shown in Table \ref{tab:The-mean-value}. }

\subsection{Coordinated Turn Motion \label{subsec:Coordinate-turning}}

The extended object undergoing coordinated turn is simulated in this
case. The extended object with the shape of the plus-sign starts from
$\left[\begin{array}{cc}
0, & 80\end{array}\right]^{T}\mathrm{cm}$ with velocity $\left[\begin{array}{cc}
5, & 1\end{array}\right]^{T}\unitfrac{cm}{s}$ at time $t=0$, then it executes a $1^\circ/s$ coordinated turn
for $60$ seconds. The extended elliptic object executes a $3^\circ/s$
coordinated turn for $60$ seconds. The sampling interval is $10$
seconds. At each time step, $100$ noisy measurement points are uniformly
generated from the extents of the targets. The noise variance is selected
as $\mathbf{diag}(\begin{array}{cc}
1, & 1\end{array})$ for each point measurement.

The extended objects executing a coordinated turn are estimated based
on the dynamic model (\ref{eq:dynamic CT-1}). The states are initialized
as a circular shape with radius of $7.8\mathrm{cm}$. The tracking
results for ellipse-shaped object are shown in Fig. \ref{fig:constant velocity model}(c)
and tracking results for plus-sign-shaped object are shown in Fig.
\ref{fig:constant velocity model}(d). \textcolor{black}{The mean
value of the RMSE of the position over $100$ Monte Carlo runs is
$0.81\mathrm{cm}$ for ellipse and $0.96\mathrm{cm}$ for the plus-sign.
The mean value of the RMSE of the velocity over $100$ Monte Carlo
runs is $0.45\unitfrac{cm}{s}$ for ellipse and $0.32\unitfrac{cm}{s}$
for the plus-sign. The mean values of the IoU of the ellipse and the
plus-sign-shaped targets during the coordinated turn motion are shown
in Table \ref{tab:The-mean-value}. }The image moments based model,
which provides a dynamic model for the shape of extended object undergoing
a coordinated turn, can estimate the positions and velocities of the
target, as well as the orientations and extents of the targets very
accurately. 

\subsection{Complex trajectories}

The image moments based RHM is embedded in the IMM framework. The
proposed model is tested in two simulations of the extended elliptical
objects switching between maneuvering and non-maneuvering intervals
multiple times.

\subsubsection{Slow motion and maneuvering case}

The target is moving with a constant velocity of $50\unitfrac{km}{h}$,
with initial state in Cartesian coordinates $\mathbf{p}_{0}=\left[x_{c},\dot{x}_{c},y_{c},\dot{y}_{c}\right]^{T}=[0,9.8,0,-9.8]^{T}$(with
position in m). The target first executes a $45^{\circ}$ coordinated
turn with the turning rate of $0.46\nicefrac{\circ}{s}$ at 260 second
for 100 seconds, then it goes through two $90^{\circ}$ coordinated
turns with the turning rate of $0.90\nicefrac{\circ}{s}$ at 570 second
and 830 second for $100$ seconds. The trajectory is shown in Fig.
\ref{fig:The-trajectory,-measurements}. The major and minor axes
of the elliptical target are set to $340$m and $80$m, respectively.
The number of the measurements in each scan is generated based on
the Poisson distribution with mean of $10$, and the measurement points
are uniformly distributed. The variance of the measurement noise is
$\mathbf{diag}(\begin{array}{cc}
10^{2}, & 10^{2}\end{array})$, and the sampling time is $10$s. 

\begin{figure}
\centering{}\includegraphics[width=0.8\columnwidth]{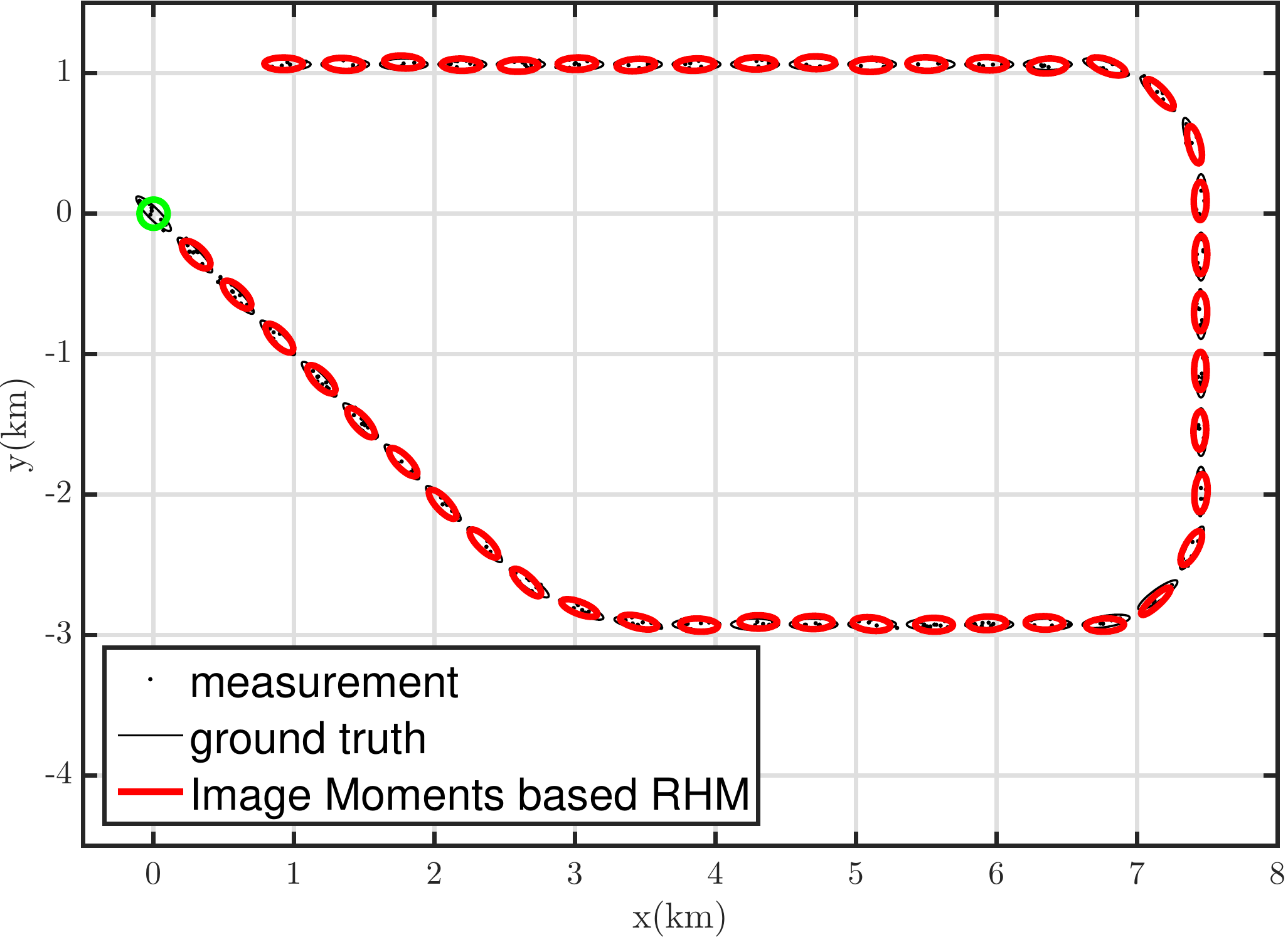}\caption{The trajectory, measurements and one example run of the simulation.
Estimation results are shown for every $30$ seconds.\label{fig:The-trajectory,-measurements}}
\end{figure}

\begin{figure*}
\begin{centering}
\makebox[0.8\linewidth][c]{\subfigure[Average RMSE of position]{\includegraphics[width=0.23\paperwidth]{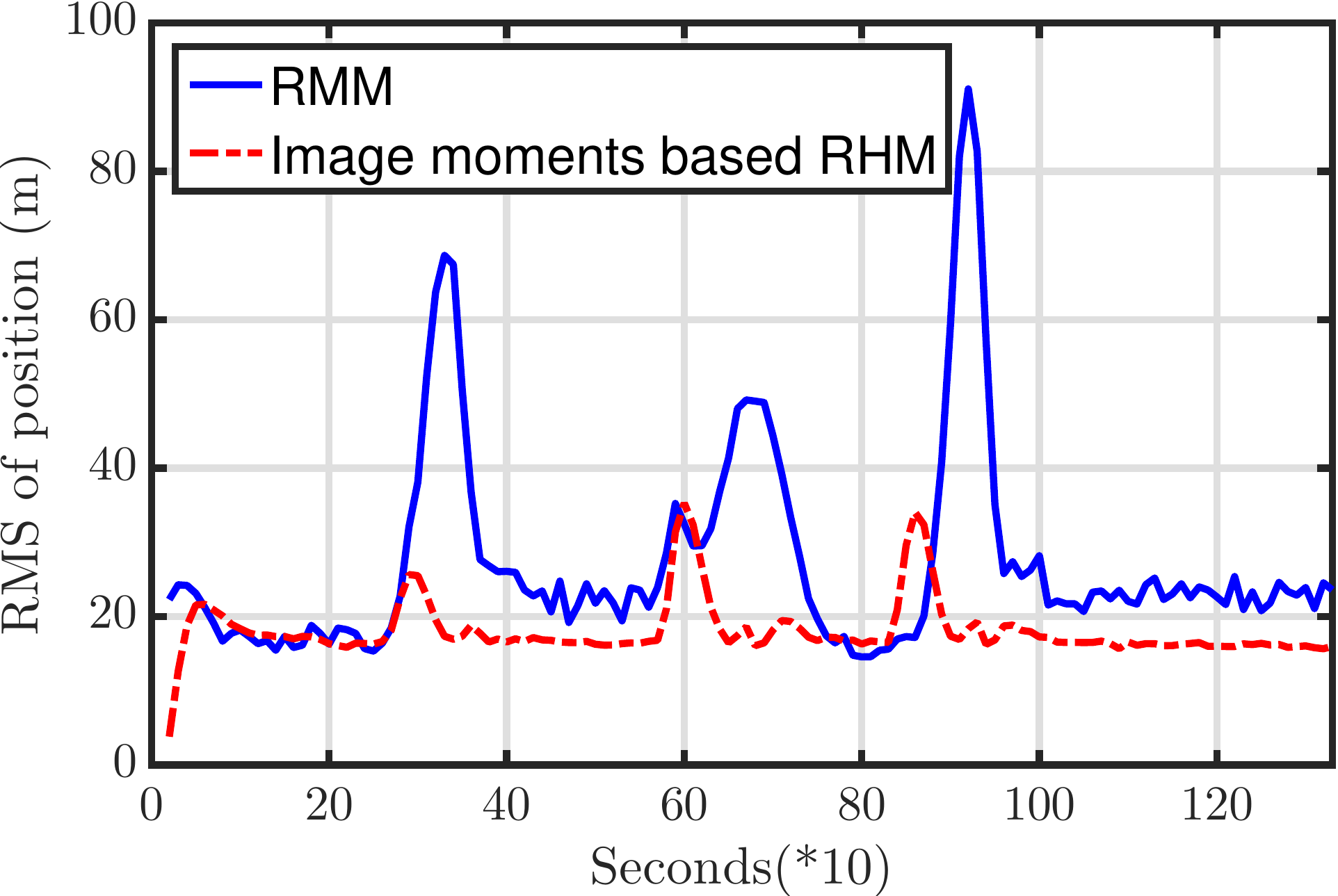}}\hspace{1em}\subfigure[Average RMSE of velocity]{\includegraphics[width=0.23\paperwidth]{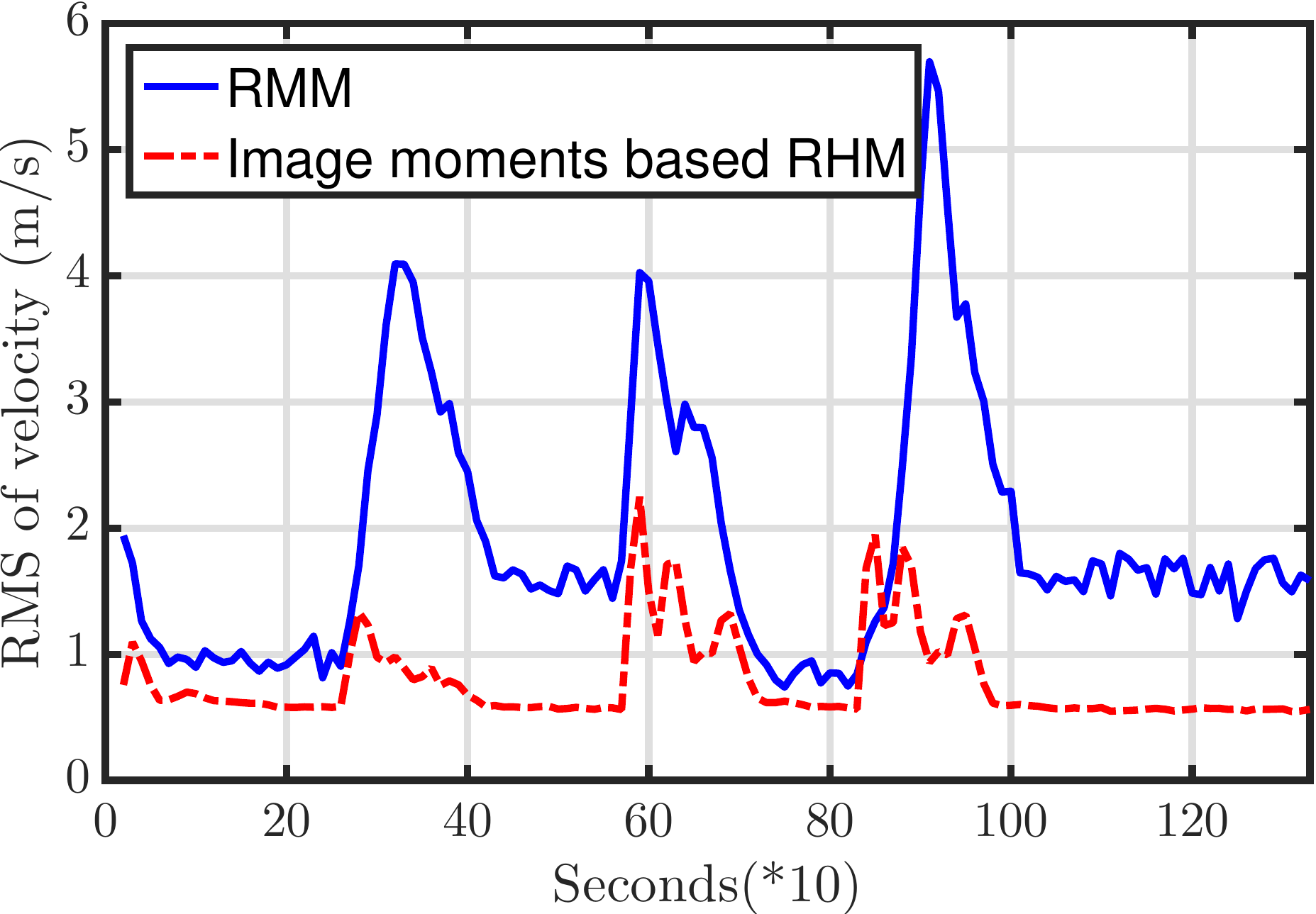}}\hspace{1em}\subfigure[Average IoU]{\includegraphics[width=0.23\paperwidth]{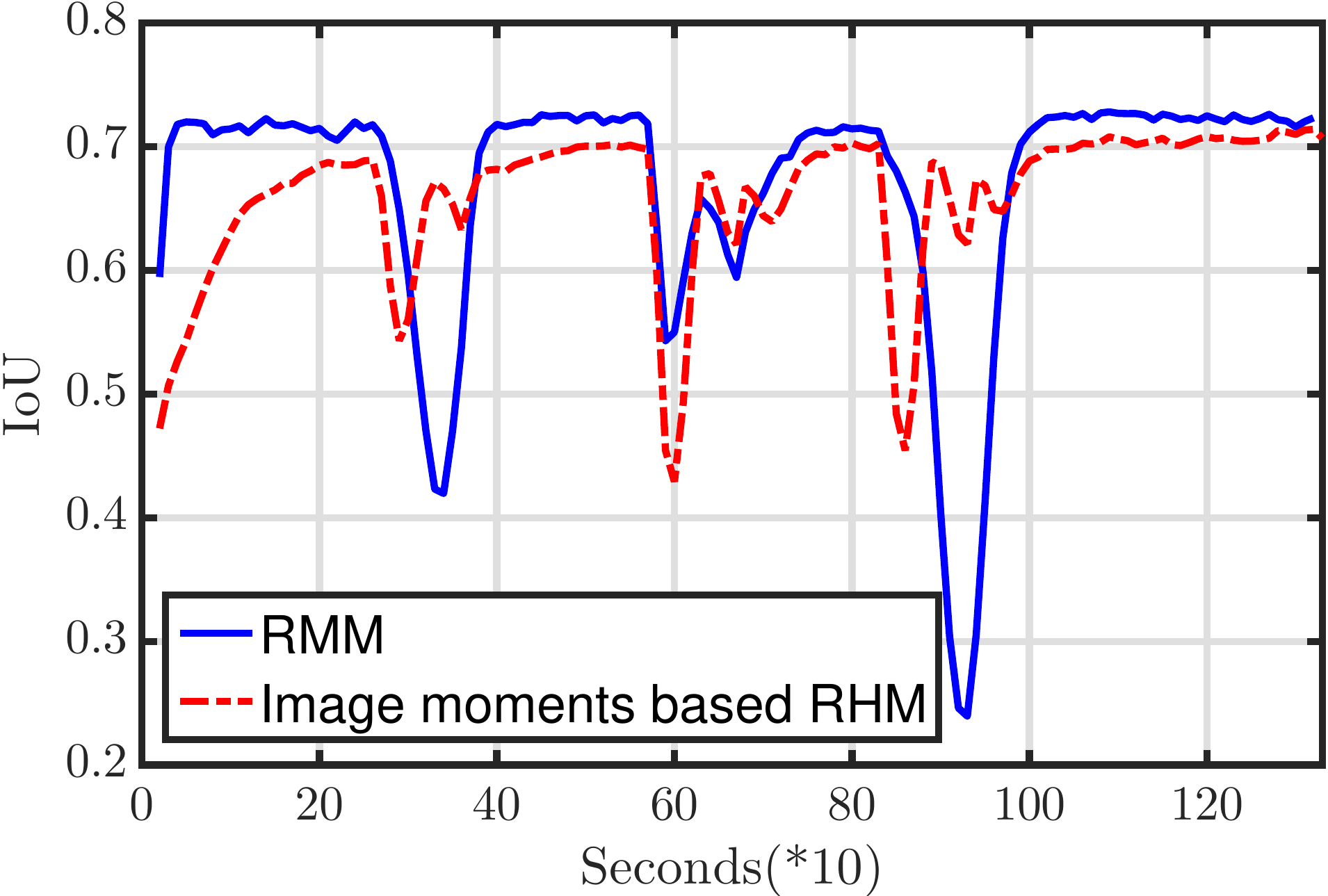}}}\\
\par\end{centering}
\centering{}\caption{Simulation results of the proposed image moments based RHM algorithm
compared with the RMM algorithm in \cite{feldmann2011tracking} over
$1000$ Monte Carlo runs: (a) The average RMSE of the position of
the centroid; (b) The average RMSE of the velocity of the centroid;
(c) The average IoU. \label{fig:Simulation-results}}
\end{figure*}

\begin{figure}
\centering{}\includegraphics[width=0.8\columnwidth]{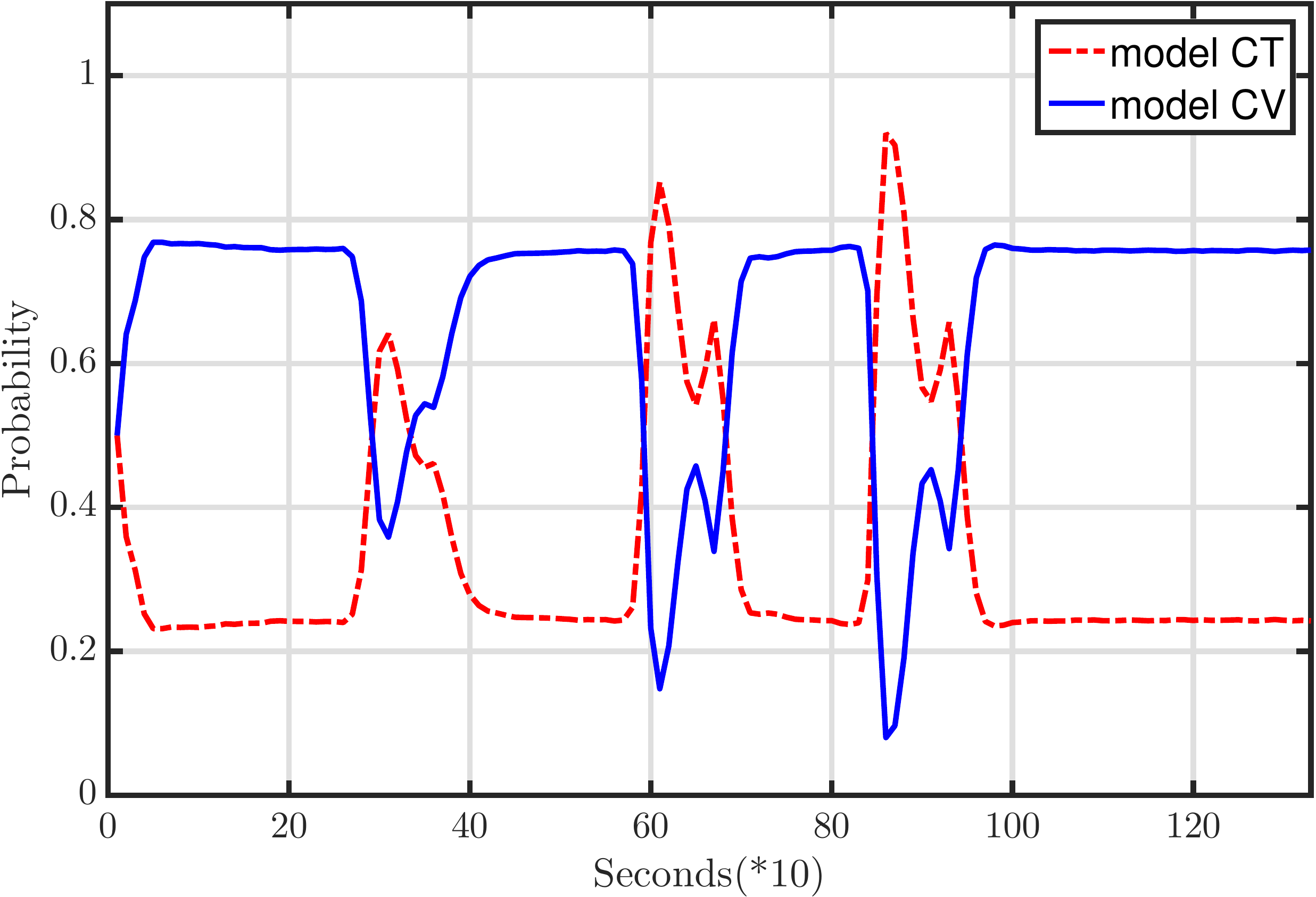}\caption{Model probability of the UKF-IMM filter for the image moments based
RHM.\label{fig:Model-probability}}
\end{figure}

The proposed Image moments based random hypersurface model with UKF-IMM
algorithm is compared with the RMM with IMM algorithm in \cite{feldmann2011tracking}.
The RMM-IMM algorithm uses two models. The model with a high kinematic
process noise and a high extension agility accounts for abrupt changes
in shape and orientation during maneuvers, and another model with
low kinematic noise and a low extension agility accounts for the non-maneuvers.
The extension agility is set as 10 and 5 separately for both models.
The kinematic states of both models use the constant velocity model
(the kinematic dynamic model in (\ref{eq:cvm-1})), and the parameter
$q$ in (\ref{eq:cvm-1}) is set as $10$ and $0.1$ respectively.
The proposed image moments based RHM with the UKF-IMM filter combines
the constant velocity model in (\ref{eq:cvm-1}) and the coordinated
turn model in (\ref{eq:dynamic CT-1}). The parameter $q$ for the
process noise covariance in the constant velocity model in (\ref{eq:cvm-1})
is set as $0.01$ and $\mathbf{C}_{\mathrm{IM}}=\mathrm{\mathbf{diag}}(1,1,1)$.
For the coordinated turn model in (\ref{eq:dynamic CT-1}), $\mathbf{w}_{k}=[0.01,0.01,0.01,0.0001,0.0001,\left(0.02\times\pi/180\right)^{2}]^{T}$.
The initial probability $\mu_{0}^{j}$ of the two models in the IMM
filter for both algorithms is set as equal and the Markov chain transition
matrix is selected to be $p_{i|j}=\left[\begin{smallmatrix}0.90 & 0.10\\
0.25 & 0.75
\end{smallmatrix}\right]$. The model probability of the proposed algorithm is shown in Fig.
\ref{fig:Model-probability}. With the same trajectory, the two algorithms
run with 1000 Monte Carlo runs and their simulation results are shown
in Fig. \ref{fig:Simulation-results}. The proposed algorithm has
lower RMS errors both for position and velocity and the RMM algorithm
in \cite{feldmann2011tracking} has better IOU values. 

\subsubsection{Fast motion and maneuvering case}

\begin{figure}
\centering{}\includegraphics[width=0.8\columnwidth]{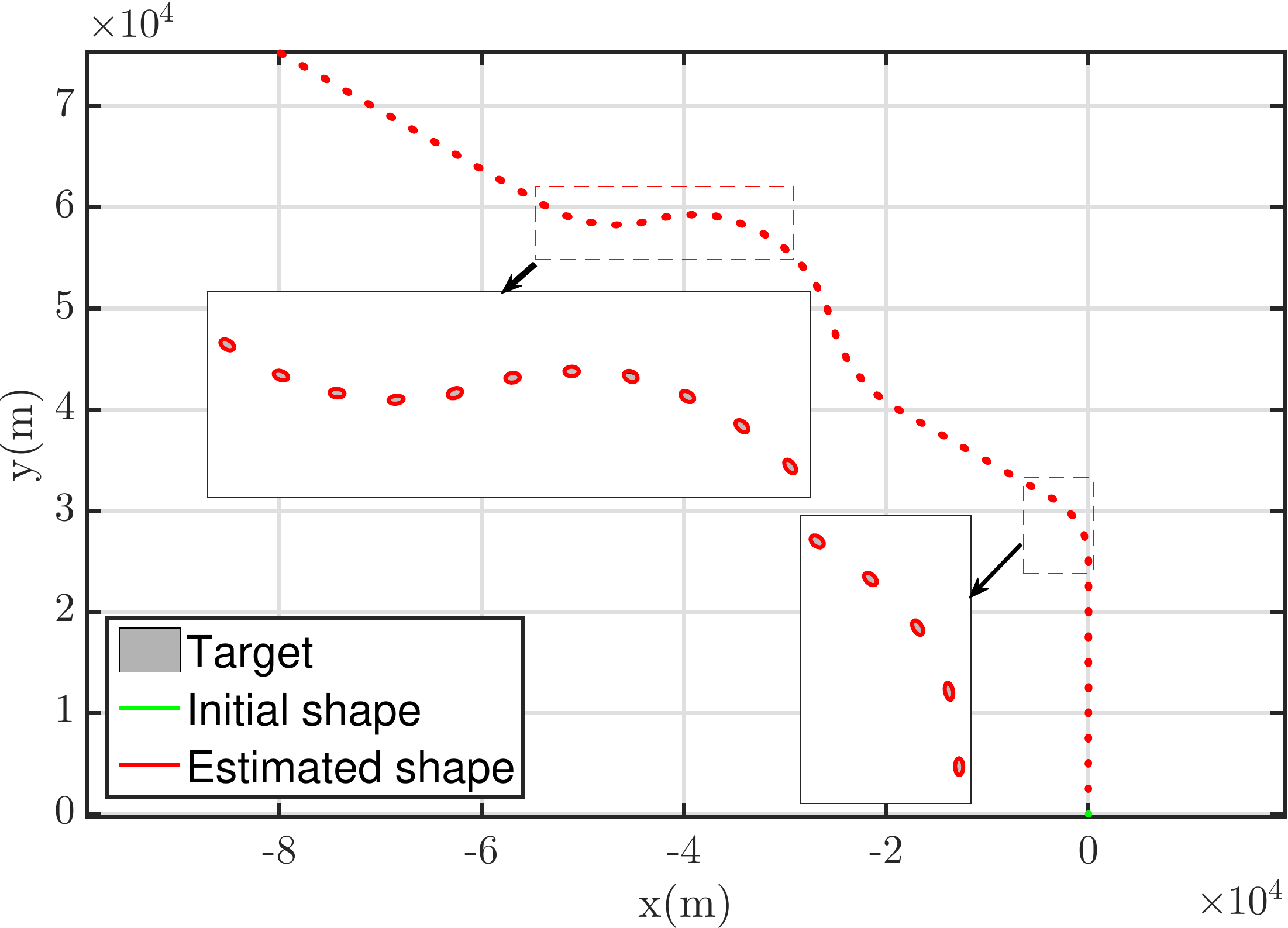}\caption{Trajectory of the extended object and the estimation results of the
image moments based RHM in one run.\label{fig:Trajectory2} }
\end{figure}

The typical trajectory from \cite{bar2004estimation} is used for
this simulation and shown in Fig. \ref{fig:Trajectory2}\textcolor{black}{.
The target is moving with a constant velocity of $250\unitfrac{m}{s}$,
with initial state in Cartesian coordinates $\mathbf{p}_{0}=\left[x_{c},\dot{x}_{c},y_{c},\dot{y}_{c}\right]^{T}=[0,0,0,250]^{T}$(with
position in m). The details of its maneuvering and non-maneuvering
intervals are shown in Table \ref{tab:IMM trajectory}. }For easily
visualizing, the major and minor axes of the elliptical target are
enlarged to $300$m and $150$m, respectively. The number of the measurements
in each scan is generated based on the Poisson distribution with mean
of $50$, and the measurement points are uniformly distributed. The
variance of the sensor noise is $\mathbf{diag}(\begin{array}{cc}
3^{2}, & 3^{2}\end{array})$, and the sampling time is $10$s. 

\begin{table}
\begin{centering}
\caption{Details on the complex trajectory. \label{tab:IMM trajectory}}
\par\end{centering}
\centering{}\resizebox{\columnwidth}{!}{%
\begin{tabular}{>{\centering}p{2cm}c>{\centering}p{2cm}cc}
\hline 
Time

(second) & Model & Turning rate

($\ensuremath{^{\circ}/s}$) & Turning direction & Acceleration\tabularnewline
\hline 
\hline 
$0-100$ & CV & 0 & $-$ & $-$\tabularnewline
\hline 
$100-130$ & CT & 2 & left & 0.89g\tabularnewline
\hline 
$130-200$ & CV & 0 & $-$ & $-$\tabularnewline
\hline 
$200-245$ & CT & 1 & right & 0.45g\tabularnewline
\hline 
$245-335$ & CT & 1 & left & 0.45g\tabularnewline
\hline 
$335-380$ & CT & 1 & right & 0.45g\tabularnewline
\hline 
$380-500$ & CV & 0 & $-$ & $-$\tabularnewline
\hline 
\end{tabular}} 
\end{table}

\begin{figure*}
\begin{centering}
\makebox[0.8\linewidth][c]{\subfigure[Average RMSE of position]{\includegraphics[width=0.25\paperwidth]{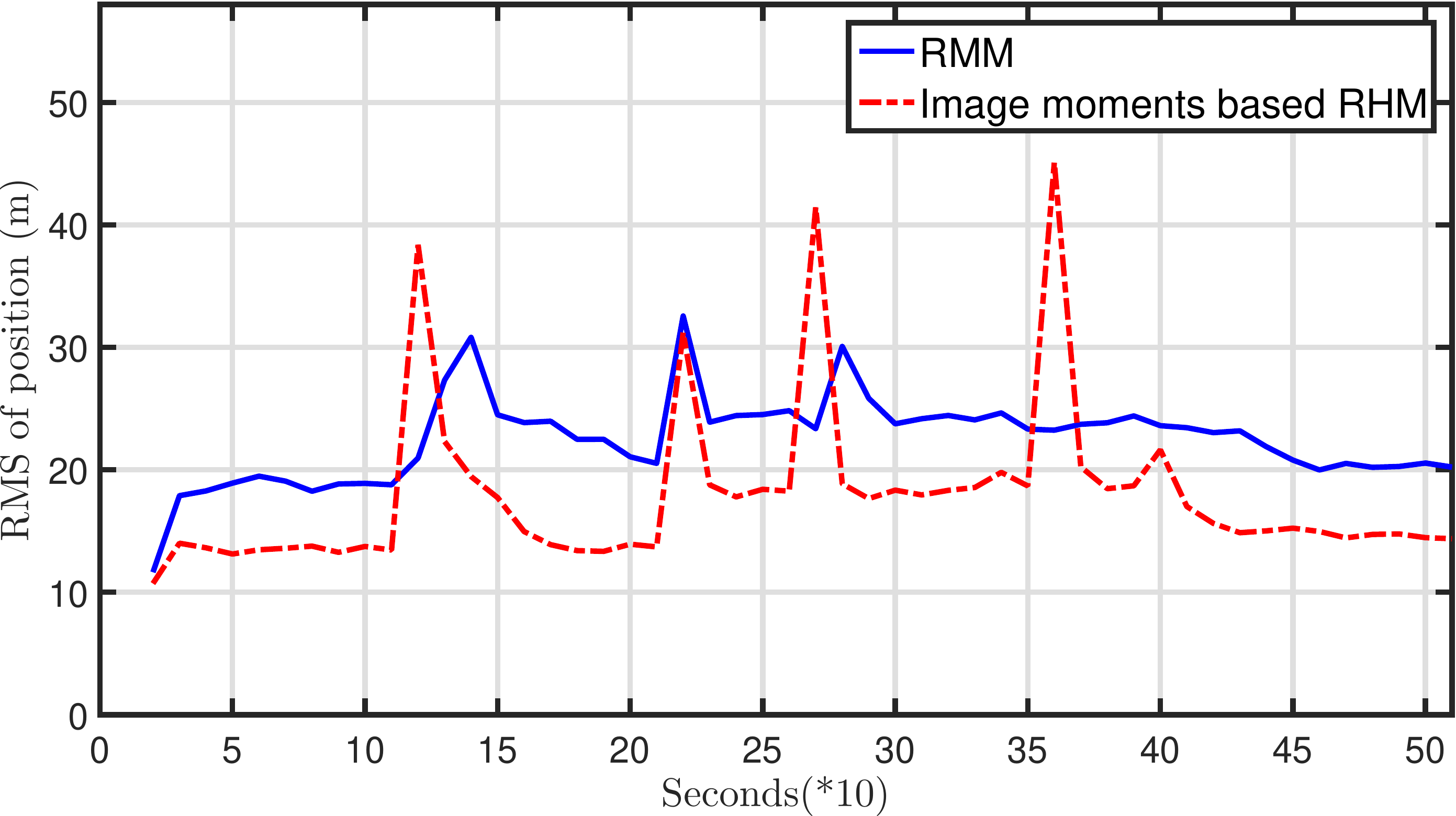}}\hspace{1em}\subfigure[Average RMSE of velocity]{\includegraphics[width=0.25\paperwidth]{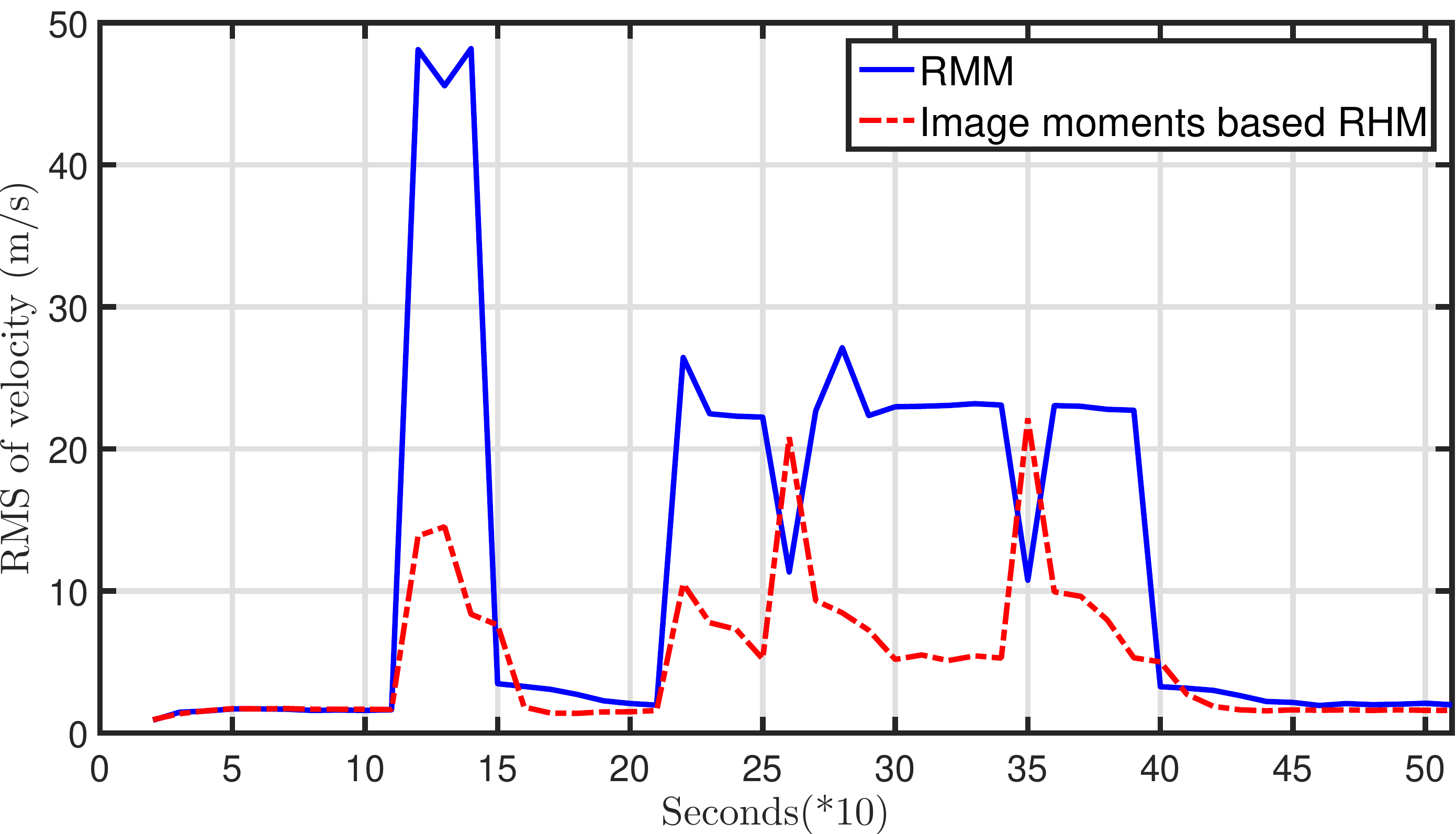}}\hspace{1em}\subfigure[Average IoU]{\includegraphics[width=0.25\paperwidth]{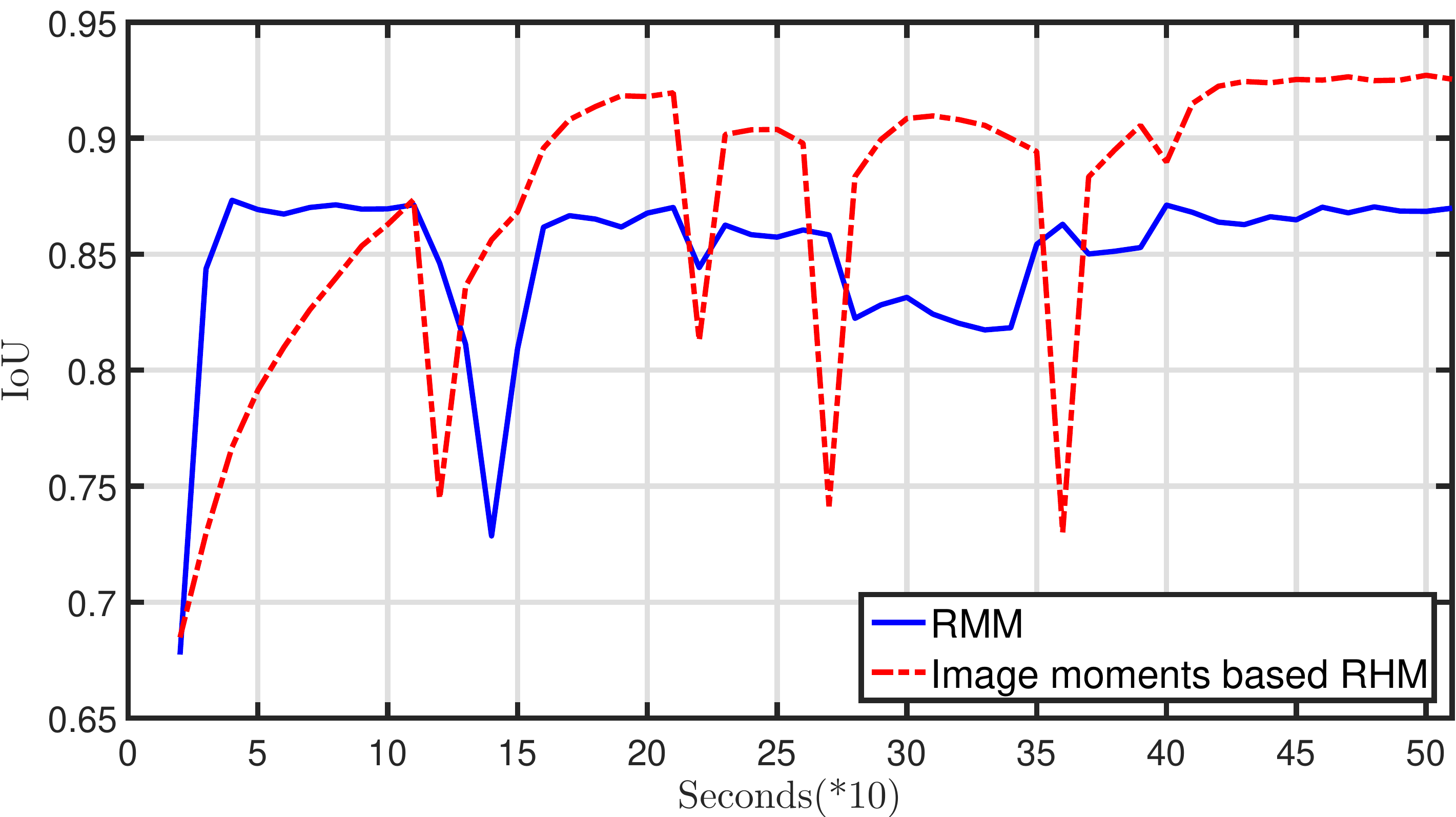}}}\\\caption{Comparison of the estimation results of the image moments based RHM
and the RMM algorithm developed in \cite{feldmann2011tracking}. \label{fig:Estimation-results2}}
\par\end{centering}
\end{figure*}

\begin{figure}
\begin{centering}
\includegraphics[width=0.8\columnwidth]{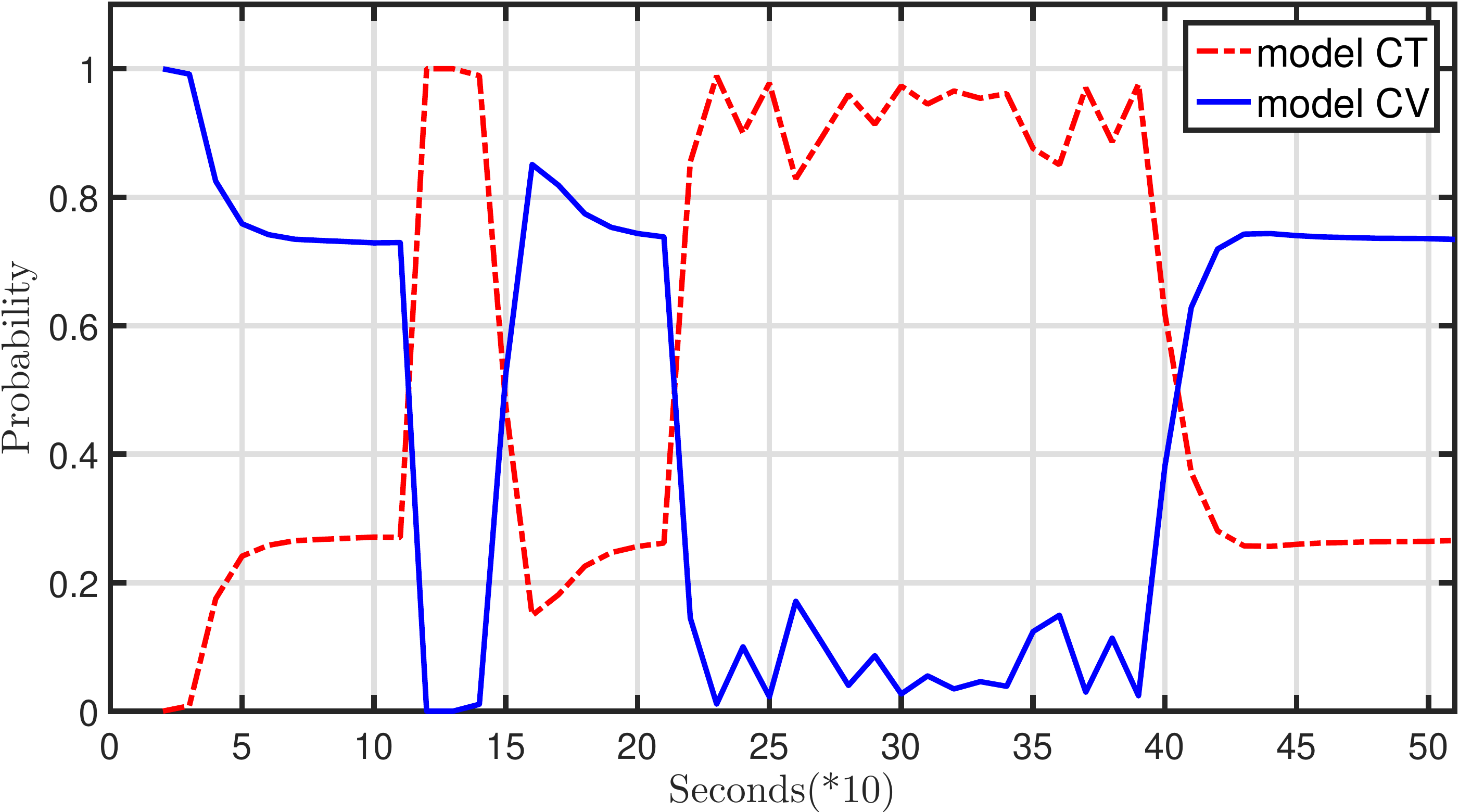}\caption{Average model probabilities in $1000$ runs.\label{fig:Average-model-probabilities2}}
\par\end{centering}
\end{figure}

The RMM-IMM algorithm \cite{feldmann2011tracking} also consists of
two models. The extension agility is set as 10 and 5 separately for
both models. The kinematic states of both models use the constant
velocity model (the kinematic dynamic model in (\ref{eq:cvm-1})),
and the parameter $q$ in (\ref{eq:cvm-1}) is set as $100$ and $0.3$
separately. The proposed image moments based RHM with the UKF-IMM
filter combines the constant velocity model in (\ref{eq:cvm-1}) and
the coordinated turn model in (\ref{eq:dynamic CT-1}). The parameter
$q$ for the process noise covariance in the constant velocity model
in (\ref{eq:cvm-1}) is set as $0.05$ and $\mathbf{C}_{\mathrm{IM}}=\mathrm{\mathbf{diag}}(50000,50000,50000)$.
For the coordinated turn model in (\ref{eq:dynamic CT-1}), $\mathbf{w}_{k}=[1000,1000,1000,0.5,0.5,\left(0.05\times\pi/180\right)^{2}]^{T}$.
The initial probability $\mu_{0}^{j}$ of the two models in the IMM
filter for both algorithms is set as equal and the Markov chain transition
matrix is selected to be $p_{i|j}=\left[\begin{smallmatrix}0.85 & 0.15\\
0.90 & 0.10
\end{smallmatrix}\right]$. The model probability of the proposed algorithm is shown in Fig.
\ref{fig:Average-model-probabilities2}. With the same trajectory,
the two algorithms run with 1000 Monte Carlo runs and their simulation
results are shown in Fig. \ref{fig:Estimation-results2}.

The proposed image moments based RHM and its measurement and dynamic
models are validated in the simulations of the static targets, the
targets with linear motion and with the coordinated turn motion. As
the noise levels are increased, the size of the estimated elliptical
shape doesn't increase as the sensor noise increases. When the targets
are performing during the linear motion or the coordinated turn motion,
the proposed algorithm can predict the position and velocity of the
moving target, as well as the spatial extent and orientation of the
targets. To estimate the target moves switching between the maneuvering
and non-maneuvering intervals. the proposed image moments based RHM
is embedded with the IMM framework. The proposed average measurement
log-likelihood function can estimate the model probability accurately
and consistently. The RMSE values of the position and velocity of
the target's centroid is lower than the results from the RMM. The
state variables of the RMM is the centroid and the random matrix,
which is updated based on the mean and spread matrix of the measurement
points\cite{feldmann2011tracking}. The proposed RHM using the centroid
and the three image moments as the state variables, which is updated
based on each individual measurement point. When the number of the
measurement points is small or noisy, the proposed image moments based
RHM estimates the position and velocity of the centroid accurately,
while the mean of the measurement points is far away from the position
of the centroid. The accurate dynamic model has the advantage of predicting
the location of the target, especially when predict the location of
the target undergoing fast motion and the sampling frequency is relatively
low. 
\begin{figure}
\centering{}\includegraphics[width=0.85\columnwidth]{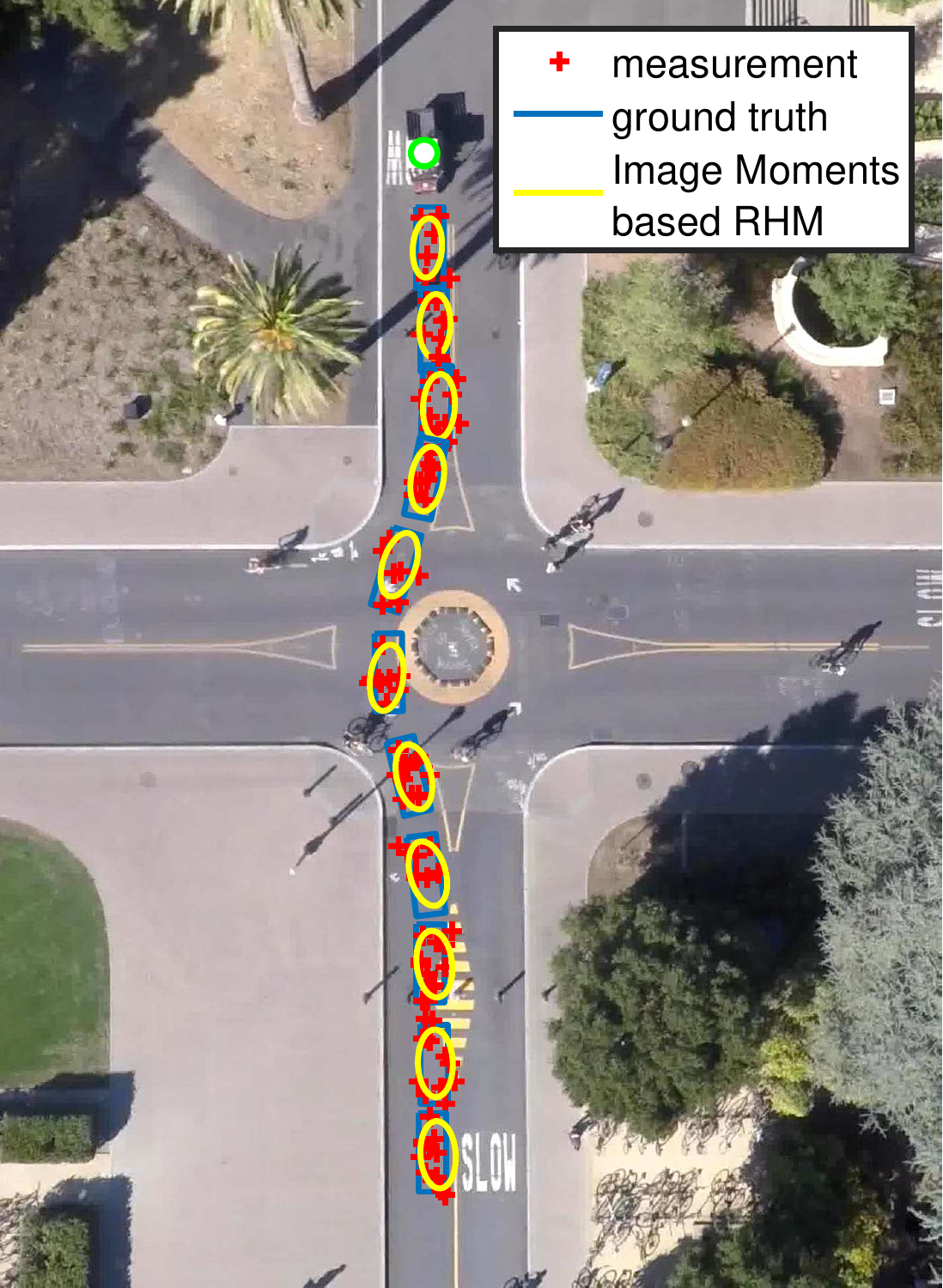}\caption{Illustration of the estimation results in a particular run by the
proposed image moments based RHM; The target is at first estimated
as the circle ( green circle ) with the radius of 20 pixels; The estimated
results (yellow ellipse), the measurements (red crossing) and the
ground truth (blue box) are shown for every 40 frames.\label{fig:car}}
\end{figure}
\begin{figure}
\begin{centering}
\makebox[0.8\linewidth][c]{\subfigure[Average RMSE of the position]{\includegraphics[width=0.9\columnwidth]{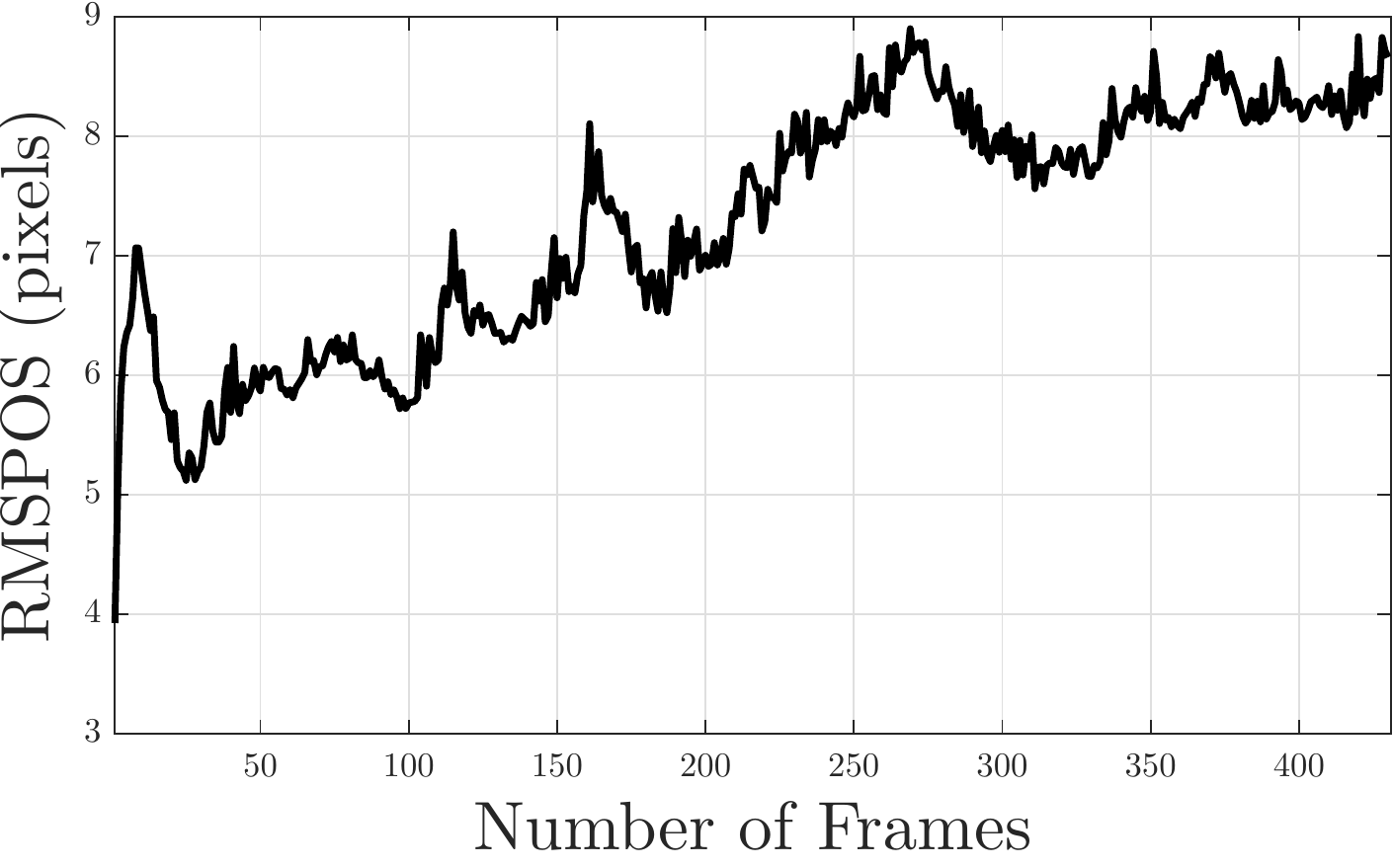}}}\\\vspace{-1.5ex}\makebox[0.8\linewidth][c]{\subfigure[Average IoU]{\includegraphics[width=0.9\columnwidth]{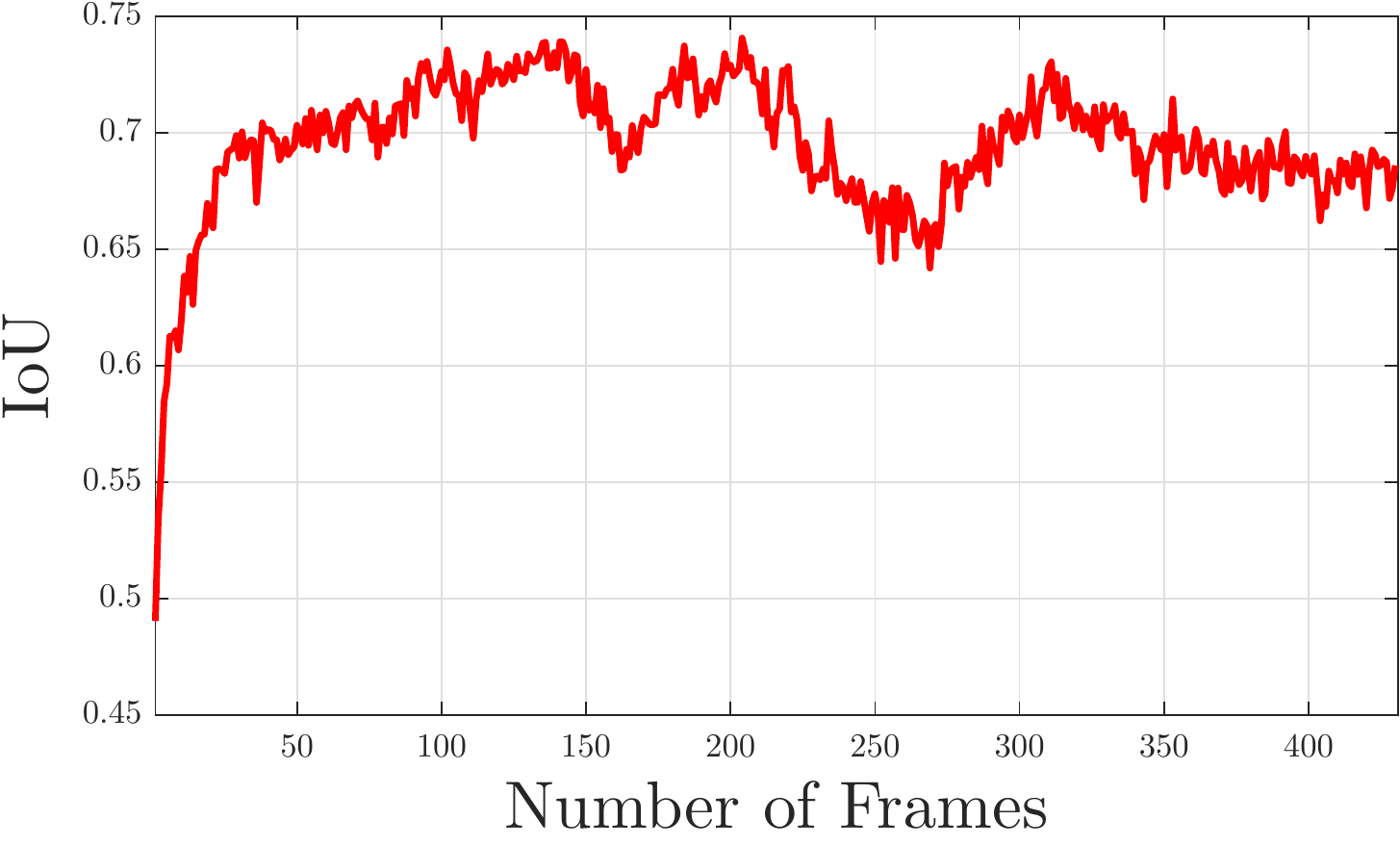}}}\\
\par\end{centering}
\caption{Tracking results of the proposed image moments based RHM algorithm
over $1000$ Monte Carlo runs. \label{fig:Tracking-results-of-car}}
\end{figure}

\section{Experiment\label{sec:Experiment}}

In this section, the proposed image moments based RHM is applied for
tracking a moving car represented as an extended object in a real
video. A short video clip from the Stanford drone dataset \cite{robicquet2016learning}
is used, which shows a moving car from a bird's eye view. The video
is captured with a 4k camera mounted on a quadcopter platform (a 3DR
solo) hovering above an intersection on a university campus at an
altitude of approximately $80$ meters which contains 431 frames with
the image size of 1422 by 1945 pixels and the video has been undistorted
and stabilized \cite{robicquet2016learning}. The ground truth is
manually labeled at each frame and the measurement points are uniformly
generated inside the bounding box of the ground truth. The number
of measurements in each frame is generated based on the Poisson distribution
with mean of $10$. The sensor noise is Gaussian white noise with
variance $\mathbf{diag}(\begin{array}{cc}
10^{2}, & 10^{2}\end{array})$. In Fig. \ref{fig:car}, the first top-view scene of the moving car
is shown and $11$ snapshots of the estimation results out of the
431 frames are plotted in the same figure. The target is moving switching
between the linear motions and the rotational motions. The constant
velocity model in (\ref{eq:cvm-1}) and the coordinated turn model
in (\ref{eq:dynamic CT-1}) with the UKF-IMM filter are applied to
track the moving car. The parameter $q$ for the process noise covariance
in the constant velocity model in (\ref{eq:cvm-1}) is set as $0.1$
and $\mathbf{C}_{\mathrm{IM}}=\mathrm{\mathbf{diag}}(0.01,0.01,0.01)$.
For the coordinated turn model in (\ref{eq:dynamic CT-1}), $\mathbf{w}_{k}=[0.01,0.01,0.01,0.01,0.01,\left(0.1\times\pi/180\right)^{2}]^{T}$
(with position is in pixels). The initial probability $\mu_{0}^{j}$
of the two models in the IMM filter for both algorithms is set as
equal and the Markov chain transition matrix is selected to be $p_{i|j}=\left[\begin{smallmatrix}0.90 & 0.10\\
0.10 & 0.90
\end{smallmatrix}\right]$.

The proposed algorithms run with 1000 Monte Carlo runs and their estimation
results are shown\textcolor{black}{{} in Fig. \ref{fig:Tracking-results-of-car}.
The mean value of the RMSE of the centroid position over $1000$ Monte
Carlo runs is $7.28\mathrm{pixels}$. The mean value of the IoU (the
ground truth is approximated as an ellipse with the axes are same
as the width and height of the corresponding bounding box and they
have the same orientation) over $1000$ Monte Carlo runs is $0.70$. }

\section{Conclusion\label{sec:Conclusion-and-Future}}

In this paper, the minimal, complete, and non-ambiguous representation
of an elliptic object is modeled based on image moments for extended
object tracking. The measurement model and the dynamic models of the
image moments for linear motion and coordinated turn motion are analytically
derived. The unscented Kalman filter and its combination with the
interacting multiple model approach is applied for estimating the
position, velocity and spatial extent based on the noisy measurement
points uniformly generated from the extended target. The proposed
image moments based random hypersurface model and its filters are
validated and evaluated in different simulation scenarios and one
real trajectory. The evaluation results show that the proposed model
and its inference can provide accurate estimations of the position,
velocity and extents of the targets. The proposed Image moments based
RHM for tracking the extended objects can be embedded into other Bayesian
based methods, such as multiple hypothesis tracking techniques or
probabilistic data association filters. 

\appendices{}

\section{Transition matrix of the coordinated turn motion\label{sec:appendix A}}

\begin{equation}
\mathbf{\dot{p}}_{\mathrm{IM}}=\mathbf{A}\mathbf{p}_{\mathrm{IM}}\label{eq:appendix state space}
\end{equation}
where $\mathbf{A}=\left[\begin{array}{ccc}
0 & \omega & -\omega\\
-2\omega & 0 & 0\\
2\omega & 0 & 0
\end{array}\right]$. The solution to this linear time-invariant state space equation
(\ref{eq:appendix state space}) is
\begin{equation}
\mathbf{p}_{\mathrm{\mathrm{IM}}}(t)=e^{\mathbf{A}\tau}\mathbf{p}_{\mathrm{IM}}(t_{0})
\end{equation}
which $\tau=t-t_{0}$. 

The interpolation polynomial method \cite{bar2004estimation} is used
to get the transition matrix of the dynamic equation $f(\lambda)=e^{\lambda\tau}$.
Firstly, By solving $\left|\lambda\mathbf{I}-\mathbf{A}\right|=\lambda(\lambda^{2}+4\omega^{2})=0$,
the eigenvalues of the matrix $\mathbf{A}$ is calculated as $\lambda_{1}=0$,
$\lambda_{2}=2\omega j$ and $\lambda_{3}=-2\omega j$. Then, a polynomial
of degree of $2$ as $g(\lambda)=\sum_{k=0}^{2}g_{k}\lambda^{k}$
is found, which is equal to $f(\lambda)=e^{\lambda\tau}$ on the spectrum
of $\mathbf{A}$, that is
\begin{equation}
\frac{\partial^{j}}{\partial\lambda^{j}}g(\lambda)|_{\lambda=\lambda_{i}}=\frac{\partial^{j}}{\partial\lambda^{j}}f(\lambda)|_{\lambda=\lambda_{i}}
\end{equation}
 which $i=1,\cdots,3$ and $j=0$. The polynomial $g(\lambda)$ is
calculated as
\begin{equation}
g(\lambda)=1+\frac{sin(2\omega\tau)}{2\omega}\lambda+\frac{sin^{2}(\omega\tau)}{2\omega^{2}}\lambda^{2}
\end{equation}
Then, $f(\mathbf{A})=e^{\mathbf{A}\tau}$ is calculated by making
it equal to $g(\mathbf{A})$. The transition matrix $f(\mathbf{A})=e^{\mathbf{A}\tau}$
is calculated as $e^{\mathbf{A}\tau} = \left[\begin{smallmatrix} \mathrm{cos}2\theta & \frac{1}{2}\mathrm{sin}2\theta & -\frac{1}{2}\mathrm{sin}2\theta \\ -\mathrm{sin}2\theta & \mathrm{cos}^2\theta & \mathrm{sin}^2\theta \\ \mathrm{sin}2\theta & \mathrm{sin}^2\theta & \mathrm{cos}^2\theta\end{smallmatrix} \right]$,
where $\theta=\omega\tau$.

\section{Derivation and Moment matching of the random variable $m$\label{sec:appendix B}}

Consider the real measurement $\mathbf{z}=[\begin{array}{cc}
\tilde{x}, & \tilde{y}\end{array}]^{T}$ of the unknown true measurement $\mathbf{\bar{z}}=[\begin{array}{cc}
x, & y\end{array}]^{T}$ is expressed as $\mathbf{z}=\mathbf{\bar{z}}+\mathbf{\boldsymbol{\nu}}$,
where $\mathbf{\boldsymbol{\nu}}=\left[\nu_{x},\nu_{y}\right]^{T}$is
the additive white Gaussian noise with $\nu_{x}\sim\mathcal{N}(\begin{array}{cc}
0, & \sigma_{x}^{2}\end{array})$, $\nu_{y}\sim\mathcal{N}(\begin{array}{cc}
0, & \sigma_{y}^{2}\end{array})$ and they are independent with each other.\textcolor{red}{{} }Replacing
the unknown true measurement $\mathbf{\bar{z}}$ with the real measurement
$\mathbf{\mathbf{z}=\mathbf{\bar{z}}+\mathbf{\boldsymbol{\nu}}}$
in (\ref{eq:moment_ellipse_cartesian}) and separate the terms including
the noise $\mathbf{\boldsymbol{\nu}}$ as 
\begin{equation}
g(\mathbf{z},\mathbf{p})=g(\bar{\mathbf{z}},\mathbf{p})-f(\mathbf{z},\mathbf{\boldsymbol{\nu}},\mathbf{p})
\end{equation}
where $f(\mathbf{z},\mathbf{\boldsymbol{\nu}},\mathbf{p})$ is the
polynomial containing the white noise terms as

\begin{equation}
\begin{split}f(\mathbf{z},\mathbf{\boldsymbol{\nu}},\mathbf{p})=\rho & \biggl[\nu_{x}^{2}n_{02}+\nu_{y}^{2}n_{20}+2\nu_{x}\nu_{y}n_{11}\\
 & +2(n_{02}\nu_{x}-n_{11}\nu_{y})(\widetilde{x}-x_{c})\\
 & +2(n_{20}\nu_{y}-n_{11}\nu_{x})(\widetilde{y}-y_{c})\biggr]
\end{split}
\label{eq:w-1}
\end{equation}
where $\rho=\nicefrac{1}{4\left(n_{20}n_{02}-n_{11}^{2}\right)}$.
The polynomial $f(\mathbf{z},\mathbf{\boldsymbol{\nu}},\mathbf{p})$
can be considered as a random variable with Gaussian distribution,
which has the same mean and covariance as $f(\mathbf{z},\mathbf{\boldsymbol{\nu}},\mathbf{p})$
by moment matching. The closed-form expression of the first two moments
of $f(\mathbf{z},\mathbf{\boldsymbol{\nu}},\mathbf{p})$ are
\begin{equation}
E\left[f(\mathbf{z},\mathbf{\boldsymbol{\nu}},\mathbf{p})\right]=\rho\biggl[n_{02}\sigma_{x}^{2}+n_{20}\sigma_{y}^{2}\biggr]
\end{equation}

\begin{equation}
\begin{split}E\left[f(\mathbf{z},\mathbf{\boldsymbol{\nu}},\mathbf{p})^{2}\right]=\rho^{2} & \biggl\{3n_{02}^{2}\sigma_{x}^{4}+3n_{20}^{2}\sigma_{y}^{4}+(2n_{02}n_{20}+4n_{11}^{2})\sigma_{x}^{2}\sigma_{y}^{2}\\
 & +4\left[n_{02}(x-x_{c})-n_{11}(y-y_{c})\right]^{2}\sigma_{x}^{2}\\
 & +4\left[n_{20}(y-y_{c})-n_{11}(x-x_{c})\right]^{2}\sigma_{y}^{2}\biggr\}
\end{split}
\end{equation}
The covariance of $f(\mathbf{z},\mathbf{\boldsymbol{\nu}},\mathbf{p})$
is derived as
\begin{equation}
\begin{aligned}C_{f(\mathbf{z},\mathbf{\boldsymbol{\nu}},\mathbf{p})} & =E\left[f(\mathbf{z},\mathbf{\boldsymbol{\nu}},\mathbf{p})^{2}\right]-E\left[f(\mathbf{z},\mathbf{\boldsymbol{\nu}},\mathbf{p})\right]^{2}\\
 & =\rho^{2}\biggl\{2n_{02}^{2}\sigma_{x}^{4}+2n_{20}^{2}\sigma_{y}^{4}+4n_{11}^{2}\sigma_{x}^{2}\sigma_{y}^{2}\\
 & +4\left[n_{02}(x-x_{c})-n_{11}(y-y_{c})\right]^{2}\sigma_{x}^{2}\\
 & +4\left[n_{20}(y-y_{c})-n_{11}(x-x_{c})\right]^{2}\sigma_{y}^{2}\biggr\}
\end{aligned}
\end{equation}

\bibliographystyle{IEEEtran}
\bibliography{extendedobjecttracking}

\end{document}